\documentclass[superscriptaddress, reprint, longbibliography,
 amsmath,amssymb,
aps,
prb,
]{revtex4-1}

\usepackage{graphicx}% Include figure files
\usepackage{dcolumn}% Align table columns on decimal point
\usepackage{bm}% bold math
\usepackage{adjustbox}
\usepackage[colorlinks=true,linkcolor=blue,citecolor=blue,urlcolor=blue]{hyperref}
\usepackage{braket}

\newcommand{\I}{\mathrm{Im}}
\newcommand{\FMd}{\mathrm{FMd}}
\usepackage{bbm}
\newcommand{\ex}{\mathrm{ex}}
\newcommand{\IP}{\mathrm{IP}}

\newcommand{\x}{\mathrm{x}}
\newcommand{\FM}{\mathrm{FM}}

\newcommand{\UE}{\mathrm{UE}}

\newcommand{\eh}{\mathrm{eh}}

\newcommand{\emm}{\mathrm{em}}
\newcommand{\abs}{\mathrm{abs}}

\begin{document}

%\preprint{APS/123-QED}

\title{Rearrangement collision theory of phonon-driven exciton dissociation} 

\author{Christopher J. N. Coveney}
\affiliation{Department of Physics, University of Oxford, Oxford OX1 3PJ, United Kingdom}
\author{Jonah B. Haber}
\affiliation{Materials Sciences Division, Lawrence Berkeley National Laboratory, Berkeley, California 94720, United States}
%\affiliation{Department of Physics, University of California Berkeley, Berkeley, United States}
\affiliation{Department of Materials Science and Engineering, Stanford University, Stanford, California 94305, United States}
\author{Antonios M. Alvertis}
\affiliation{Materials Sciences Division, Lawrence Berkeley National Laboratory, Berkeley, California 94720, United States}
%\affiliation{Department of Physics, University of California Berkeley, Berkeley, United States}
\affiliation{KBR, Inc, NASA Ames Research Center, Moffett Field, California 94035, United States}
\author{Jeffrey B. Neaton}
%\email{jbneaton@lbl.gov}
\affiliation{Materials Sciences Division, Lawrence Berkeley National Laboratory, Berkeley, California 94720, United States}
\affiliation{Department of Physics, University of California Berkeley, Berkeley, United States}
\affiliation{Kavli Energy NanoScience Institute at Berkeley, Berkeley, United States}
\author{Marina R. Filip}
\email{marina.filip@physics.ox.ac.uk}
\affiliation{Department of Physics, University of Oxford, Oxford OX1 3PJ, United Kingdom}

\date{\today}

\begin{abstract}
Understanding the processes governing the dissociation of excitons to free charge carriers in semiconductors and insulators is of central importance for photovoltaic applications. Dyson's $\mathcal{S}$-matrix formalism provides a framework for computing scattering rates between quasiparticle states derived from the same underlying Hamiltonian, often reducing to familiar Fermi’s golden rule like expressions at first order. By presenting a rigorous formalism for multi-channel scattering, we extend this approach to describe scattering between composite quasiparticles and in particular, the process of exciton dissociation mediated by the electron-phonon interaction. Subsequently, we derive rigorous expressions for the exciton dissociation rate, a key quantity of interest in optoelectronic materials, which enforce correct energy conservation and may be readily used in \emph{ab initio} calculations. We apply our formalism to a three-dimensional model system to compare temperature-dependent exciton rates obtained for different scattering channels.   
\end{abstract}


\maketitle

%\mathcal{T}ableofcontents

\section{Introduction}
\emph{Ab initio} calculations of the properties of quasiparticles such as electrons, polarons, excitons, trions, biexcitons and more, have proven to be a powerful means for understanding and predicting spectroscopic phenomena in molecules and materials.~\cite{hybertsen1986electron,rohlfing1998electron,giustino2017electron,szyniszewski2017binding,xie2021theory} In particular, understanding the dynamics governing these quasiparticles is crucial in determining their lifetimes and transport properties.~\cite{ponce2016epw} While scattering rates and cross-sections are  routinely  computed from \emph{ab initio} for a variety of physical mechanisms (e.g. electron-phonon,~\cite{marini2015many,ponce2016epw,chaput2019finite,brunin2020phonon} electron-electron,~\cite{da2007angular} electron-defect~\cite{lu2019efficient}), formalisms and workflows for computing interactions between multiparticle states are less explored. The dissociation of composite quasiparticles requires knowledge of potential scattering channels beyond the `elastic' scattering as described by Fermi's golden rule (FGR). A representative example is phonon-driven exciton dissociation, whereby an exciton scatters to a free electron-hole pair by interaction with phonons. 

Excitons are quasiparticles resulting from the attractive (screened) Coulomb potential experienced by excited electron-hole pairs.~\cite{benedict1998theory,rohlfing1998electron,rohlfing2000electron} Exciton formation and dissociation is of central importance for optoelectronic devices such as light-emitting diodes and solar cells, and is central to the efficiency of photochemical reactions.~\cite{shi2013exciton,herz2018lattice,bokdam2016role,ginsberg2020spatially,grancini2013hot,dimitriev2022dynamics,Bechstedt2019,schleife2018optical} Exciton dissociation to free charge carriers and radiative recombination can be induced by a strong applied electric field,~\cite{haastrup2016stark,kamban2019field} or, in the weak field regime, via exciton-phonon interactions.~\cite{perea2021phonon,Bechstedt2019,Park2022,motti2023exciton}

Early theoretical work on exciton-phonon interactions in optoelectronic materials used model effective Hamiltonians to understand temperature-dependent band renormalization and scattering lifetimes between bound exciton states.~\cite{toyozawa1958theory,toyozawa1959dynamical,mahanti1972effective,yarkony1976comments,pollmann1977effective,sumi1971urbach,piermarocchi1997exciton,gurioli1998exciton,umlauff1998direct,siantidis2001dynamics} Recently, there have been several attempts to calculate exciton-phonon interactions from first principles.~\cite{marini2008ab,antonius2022theory,chen2020exciton,cudazzo2020first,perea2021phonon,chan2023exciton,paleari2022exciton} One of the earliest schemes was based on renormalization and broadening of the band structure before solving the Bethe-Salpeter equation (BSE),~\cite{marini2008ab} neglecting the scattering between finite-momentum bound exciton states.~\cite{antonius2022theory} More recent work derived formal FGR expressions for exciton--exciton scattering from many-body perturbation theory (MBPT), enforcing the energy, momentum and spin conservation.~\cite{brem2018exciton,brem2019intrinsic,christiansen2019theory,antonius2022theory,chen2020exciton,chan2023exciton,paleari2022exciton} However, since initial and final states in this process are bound exciton states, this approach does not describe explicit exciton dissociation. In addition, previous work on scattering between different quasiparticle states describe exciton--electron dissociation in semiconductor quantum wells~\cite{ramon2003theory} and monolayer transition-metal dichalcogenides.~\cite{shahnazaryan2017exciton,yang2022relaxation} These works assume the existence of a FGR scattering channel for these processes despite the fact that the initial and final states correspond to eigenstates of different reference Hamiltonians.

Until recently,~\cite{filip2021phonon,alvertis} an \emph{ab initio} description of phonon-driven exciton scattering has been restricted to scattering between bound exciton states, not exciton dissociation to free electron-hole pairs. This is usually accounted for within the MBPT formalism which gives FGR-like expressions for scattering rates between eigenstates of the same reference Hamiltonian (\emph{i.e.}~the exciton Hamiltonian). The important technological process of the dissociation of excitons to free charge carriers is a case where the initial state (exciton) and final state (free electron-hole pair) are no longer eigenstates of the same reference Hamiltonian. The usual FGR expression for the scattering rate, in principle, is applicable only if the initial and final states are orthogonal and can be shown to be eigenstates of the same reference Hamiltonian. The quantum mechanical description of such multi-channel scattering events lies beyond the realm of single particle scattering theory and requires the theory of rearrangement collisions.~\cite{lippmann1956rearrangement,epstein1957theory,sunakawa1960theory, day1961note,mittleman1961formal,chen1966formal,hahn1967theory} In this work, we provide a general framework for quasiparticle scattering, without restriction to elastic scattering, focusing specifically on exciton dissociation in free electrons and holes driven by phonons. 

Rearrangement collisions have been studied in depth in the context of nuclear collision events.~\cite{sunakawa1960theory,lippmann1956rearrangement} For example, the scattering of a bound proton by an incident neutron can result in several scattering events, including nucleon exchange or ionization, as observed by experiment.~\cite{sunakawa1960theory} These events are all examples of rearrangement collisions, or scattering between states of different reference Hamiltonians. Within rearrangement collision theory, the scattering between eigenstates of Hamiltonians in different reduced Hilbert spaces is defined by an extended $\mathcal{S}$-matrix and modified initial and final states that are both eigenstates of the same unperturbed Hamiltonian defined in a larger Hilbert space.~\cite{lippmann1956rearrangement,sunakawa1960theory} This procedure allows one to treat scattering rigorously with the standard Dyson $\mathcal{S}$-matrix theory~\cite{dyson1949s} and generate expressions for scattering rates based on the generalized optical theorem.~\cite{sunakawa1960theory} 

Here, we apply the theory of rearrangement collisions to describe exciton scattering processes. We derive modified expressions for the final free charge carrier state to ensure that it is an eigenstate of the initial exciton Hamiltonian and orthogonal to the initial exciton state. This expression allows us to define a common scattering operator, and to further derive expressions for the rates of these processes. The orthogonalization of the final free charge carrier state with respect to the initial exciton state is a natural procedure which makes no assumption of the character of the final state. This procedure is distinct from, but similar in spirit to, the approach of Ref.~\citenum{perea2021phonon} which introduces orthogonal plane waves to describe the `continuum states' thereby avoiding `unphysical overlaps' with the bound exciton states. However, our orthogonalization procedure is more general and arises as a consequence of the rearrangement collision theory formalism. We show that the first Born approximation for the exciton--free electron-hole pair dissociation process is well defined and reduces to the expression given in Ref.~\citenum{alvertis}. We also derive a modified expression for the exciton dissociation rate based on the renormalization of the conduction and valence bands.

The paper is organized as follows. In Section~\ref{sec:FGR}, we summarize the relationship between MBPT and FGR for scattering rates between exciton states. In Section~\ref{sec:rearrange}, we use rearrangement collision theory to derive an expression for the phonon-driven exciton dissociation rate. In Section~\ref{sec:results}, we apply our formalism to calculate dissociation rates as a function of temperature for a model system, based on the Fröhlich~\cite{doi:10.1080/00018735400101213} and hydrogenic exciton~\cite{Wannier1937,TF9383400500} approximations and compare our approach with expressions for exciton $\rightarrow$ exciton and free electron-hole $\rightarrow$  free electron-hole scattering found in the literature. Finally, we summarize our findings and provide outlook for future work in Section~\ref{sec:conclusion}.  

\section{Review of Fermi's Golden Rule and Scattering between exciton states}\label{sec:FGR}

The interacting exciton-phonon Hamiltonian can be written as 
\begin{gather}
    \begin{split}\label{eq:ham}
        H = H^{\IP}_0  + V_{eh} + H_{ph} + V_{ep} \ ,
    \end{split}
\end{gather} 
where $H^{\IP}_0$ is the independent electron-hole pair Hamiltonian, $V_{eh}$ is the electron-hole interaction, $H_{ph}$ describes the lattice dynamics of the solid and $V_{ep}$ is the electron-phonon interaction. The electron-hole interaction, $V_{eh}$, consists of an attractive direct term and a repulsive exchange term.~\cite{rohlfing2000electron}  The attractive direct interaction is the frequency dependent, screened Coulomb interaction and is responsible for the formation of bound excited states. The repulsive exchange term is responsible for the exciton singlet-triplet splitting.~\cite{rohlfing2000electron} As the electron-phonon interaction does not couple to the spin degrees of freedom, it is spin diagonal and will therefore not act to induce transitions between different exciton spin states. Therefore, in what follows, we suppress spin indices associated with excitonic or free electron-hole states.

The exciton Hamiltonian is given by the sum of the independent electron-hole contribution and the electron-hole interaction in Eq.~\ref{eq:ham}, $H_{\ex} = H^{\IP}_0 + V_{eh}$. Written in the electron-hole basis, within the quasi-boson and Tamm-Dancoff approximations (TDA), the clamped-ion BSE for a solid is given in reciprocal space by~\cite{rohlfing2000electron,haber2023maximally}
\begin{gather}
	\begin{split}
	 &(\epsilon_{c\mathbf{k+Q}}-\epsilon_{v\mathbf{k}})A^{S\mathbf{Q}}_{vc\mathbf{k}}  \\
  &+ \sum_{v'c'\mathbf{k'}}  \braket{vc\mathbf{k}\mathbf{Q}|V_{eh}|v'c'\mathbf{k'Q}}A^{S\mathbf{Q}}_{v'c'\mathbf{k'}} 
  = \Omega_{S\mathbf{Q}}A^{S\mathbf{Q}}_{vc\mathbf{k}},
	\end{split}
\end{gather}
where $\epsilon_{c\mathbf{k+Q}}-\epsilon_{v\mathbf{k}}$ are the eigenvalues of $H^{\IP}_0$ and correspond to the conduction $(c)$ and valence $(v)$ band single-particle electronic states, respectively. We employ the TDA for simplicity, although our derivation may be readily extended beyond this approximation. The exciton eigenvalues are given by $\Omega_{S\mathbf{Q}}$, with $A^{S\mathbf{Q}}_{vc\mathbf{k}}$ corresponding to the coefficients of the exciton wavefunction with principle quantum number $S$ and centre-of-mass momentum $\mathbf{Q}$, as written in the free electron-hole basis~\cite{haber2023maximally}
\begin{equation}\label{eq:tda}
    \Psi_{S\mathbf{Q}}(\mathbf{r}_e,\mathbf{r}_h) =\sum_{vc\mathbf{k}} A^{S\mathbf{Q}}_{vc\mathbf{k}} \psi_{c\mathbf{k+Q}}(\mathbf{r}_e)\psi^*_{v\mathbf{k}}(\mathbf{r}_h) \ ,
\end{equation}
where $\psi_{n\mathbf{k}}(\mathbf{r})=e^{i\mathbf{k}\cdot\mathbf{r}}u_{n\mathbf{k}}(\mathbf{r})$ denotes the single-particle Bloch state and $\mathbf{r}_e/\mathbf{r}_h$ denote the electron/hole coordinates, respectively. The exciton Bloch state in real space is given by $\Psi_{S\mathbf{Q}}(\mathbf{r}_e,\mathbf{r}_h)=\braket{\mathbf{r}_e\mathbf{r}_h|S\mathbf{Q}}$, with $\psi_{c\mathbf{k+Q}}(\mathbf{r}_e)\psi^*_{v\mathbf{k}}(\mathbf{r}_h)=\braket{\mathbf{r}_e\mathbf{r}_h|vc\mathbf{k}\mathbf{Q}}$. Within the harmonic and adiabatic approximations, $H_{ph}$ can be written in second quantization as~\cite{giustino2017electron}
\begin{equation}
    H_{ph} = \sum_{\mathbf{q}\nu} \omega_{\mathbf{q}\nu} a^\dag_{\mathbf{q}\nu}a_{\mathbf{q}\nu} \ , 
\end{equation}
where $\omega_{\mathbf{q}\nu}$ are the normal mode frequencies of the solid and $a^\dag_{\mathbf{q}\nu}/a_{\mathbf{q}\nu}$ 
are the creation/annihilation operators corresponding to a phonon of wavevector $\mathbf{q}$ and branch $\nu$.~\cite{Quantum}

In the limit of a linear coupling, the first-order electron-phonon interaction can be written in the exciton basis as~\cite{antonius2022theory} 
\begin{equation}
    V_{ep} = \sum_{\substack{S'S\mathbf{Q}\\ \mathbf{q}\nu}} G_{S'S\nu}(\mathbf{Q},\mathbf{q}) A_{\mathbf{q}\nu} O^\dag_{S'\mathbf{Q+q}}O_{S\mathbf{Q}} \ ,
\end{equation}
where $A_{\mathbf{q}\nu} = a_{\mathbf{q}\nu}+a^\dag_{-\mathbf{q}\nu}$ is the phonon displacement operator and $O^\dag_{S\mathbf{Q}}/O_{S\mathbf{Q}}$
are the creation/annihilation operators for excitons in the Bloch state $\ket{S\mathbf{Q}}$. The matrix element is given by $G_{S'S\nu}(\mathbf{Q},\mathbf{q})=\braket{S'\mathbf{Q+q}|g_{\mathbf{q}\nu}|S\mathbf{Q}}$~\cite{antonius2022theory} and is explicitly written in the free electron-hole basis as~\cite{antonius2022theory,chen2020exciton,cohen2023phonon,chan2023exciton}
\begin{gather}
    \begin{split}
        G_{S'S\nu}(\mathbf{Q},\mathbf{q}) = &\sum_{vcc'\mathbf{k}} \left(A^{S'\mathbf{Q+q}}_{vc\mathbf{k}}\right)^* A^{S\mathbf{Q}}_{vc'\mathbf{k}}g_{cc'\mathbf{\nu}}(\mathbf{k+Q},\mathbf{q})\\
        -&\sum_{vv'c\mathbf{k}}\left(A^{S'\mathbf{Q+q}}_{vc\mathbf{k}}\right)^* A^{S\mathbf{Q}}_{v'c\mathbf{k+q}}g_{v'v\mathbf{\nu}}(\mathbf{k},\mathbf{q}) \ ,
    \end{split}
\end{gather}
where $g_{mn,\nu}(\mathbf{k},\mathbf{q}) = \braket{m\mathbf{k+q}|g_{\mathbf{q\nu}}|n\mathbf{k}}$ is the electron-phonon matrix element.~\cite{giustino2017electron}  The phonon modes and frequencies are usually obtained from density functional perturbation theory (DFPT) by construction and subsequent diagonalization of the dynamical matrix.~\cite{giustino2017electron,gonze1997dynamical,baroni2001phonons} The electron-phonon matrix elements may also be obtained from DFPT by computing the change in the self-consistent potential due to the phonon mode displacement.~\cite{baroni2001phonons,antonius2022theory,ponce2016epw,giustino2017electron} 

%The Hamiltonian in Eq.~\ref{eq:ham} is an operator identity, independent of the underlying `single-particle' basis introduced. Depending on the way the terms in Eq.~\ref{eq:ham} are grouped, we can define a reference Hamiltonian and a perturbation common to different scattering channels. In the Dyson $\mathcal{S}$-matrix theory, the perturbation scatters an initial eigenstate to a final eigenstate of the same reference Hamiltonian~\cite{Peskin1995,albers1976normal}. In the subSections below we will show how this approach leads to the expressions for the inverse exciton lifetime for different scattering channels previously reported in the literature~\cite{marini2008ab,antonius2022theory}.

\subsection{Phonon-driven exciton--exciton scattering}\label{sub:ex-ex}
To describe scattering between different exciton states we must first define the reference Hamiltonian, the initial and final exciton eigenstates, and the perturbation which induces the scattering between these states. Here, we focus on the \emph{physical} single-phonon emission and absorption scattering channels, whereby an exciton spontaneously emits/absorbs a phonon and is scattered into a different final exciton state. In the following, the wavevector and spin dependence of all quantities will be omitted and atomic units used throughout for notational clarity. All expressions below can be generalized by reintroducing the wavevectors under conservation of crystal momentum.

To describe exciton--exciton scattering, the Hamiltonian in Eq.~\ref{eq:ham} is partitioned as follows 
\begin{subequations}\label{eq:initial}
    \begin{equation}
        H_0 = H_{ph} + H^\IP_0 + V_{eh}
    \end{equation}
and
    \begin{equation}
        V_0 = V_{ep} \ ,
    \end{equation}
\end{subequations}
where $H_0$ is the reference Hamiltonian and $V_0$ is the perturbation that induces scattering. The eigenstates of the reference Hamiltonian, $H_0$, are exciton-phonon product states 
\begin{gather}
\begin{split}
     H_0\ket{S,n_\nu} &= E^{S,\nu}_i\ket{S,n_\nu}\\
\end{split}
\end{gather}
with energies
\begin{equation}
    E^{S,\nu}_i = \Omega_S + n_\nu\omega_\nu \ ,
\end{equation}
where $n_\nu$ denotes the number of phonons in mode $\nu$. The initial state is given by the product state of excitons in a bath of phonons of mode, $\nu$. The final exciton state after emission of a phonon, is an eigenstate of the same reference Hamiltonian, but with eigenvalue $E^{S',\nu}_{f}=\Omega_{S'}+(n_\nu+1)\omega_\nu$. The $\mathcal{S}$-matrix for this process is written as 
\begin{equation}\label{eq:norm_s}
    \mathbf{S}_{f,i} = \braket{\psi^-_{f}|\psi^+_i} 
\end{equation}
where $\ket{\psi^{\pm}_{i/f}}$ are the initial/final scattering-state wavefunctions based on the exciton--phonon product states and constructed from the following Lippmann-Schwinger equations~\cite{lippmann1950variational}
\begin{subequations}
    \begin{equation}~\label{eq:lipp}
        \ket{\psi^{\pm}_{i}} = \ket{S,n_\nu} + G^{\pm}(E^{S,\nu}_{i}) V_{ep} \ket{S,n_\nu}
    \end{equation}
and
    \begin{equation}~\label{eq:lipp1}
         \ket{\psi^{\pm}_{f}} = \ket{S',n_\nu+1} + G^{\pm}(E^{S',\nu}_{f}) V_{ep} \ket{S',n_\nu+1} ,
    \end{equation}
\end{subequations}
with $G^{\pm}(E) = (E-H\pm i\eta)^{-1}$, where $H$ is the full Hamiltonian defined in Eq.~\ref{eq:ham}. Importantly, the $\mathcal{S}$-matrix must be unitary as it contains the transformation between complete \emph{orthonormal} sets.~\cite{sachs1987physics} The $\mathcal{S}$-matrix represents the probability amplitude associated with the process of finding the system in a final state induced by a perturbation on the initial state. Eqs~\ref{eq:lipp} and~\ref{eq:lipp1} are geometric Dyson series for the initial and final scattered states that include the effects of the full perturbation $V_{ep}$ to infinite-order. As a result, the initial and final scattered states contain superpositions of states involving multiple \emph{virtual} excitations of the phonon quanta and exciton states. We restrict our analysis to the physical process of single-phonon emission/absorption due to the separability of the different physical scattering processes that occur. However, the formalism presented here may in principle also be extended to treat multiple-phonon scattering processes.

From Eqs~\ref{eq:lipp} and~\ref{eq:lipp1}, the $\mathcal{S}$-matrix for the process may be rewritten as
\begin{equation}~\label{eq:matrix}
    \mathbf{S}_{f,i} = \braket{S',n_\nu+1|\mathcal{S}(V_{ep})|S,n_\nu} , 
\end{equation}
where $\mathcal{S}(V_{ep})$ is the scattering operator (in the interaction picture) 
\begin{equation}~\label{eq:ins_s}
    \mathcal{S}(V_{ep}) = \mathcal{T}\left[\exp\left(-i\int^\infty_{-\infty}dt\ V_{ep}(t)\right)\right] ,
\end{equation}
and $\mathcal{T}$ is the time-ordering operator. Expansion of the $\mathcal{S}$-matrix in Eq.~\ref{eq:matrix}, within the Born approximation (see Appendix~\ref{app:ex_ex}), gives the exciton--exciton scattering rate (for the emission pathway):
\begin{equation}~\label{eq:ex_ex}
     \gamma^{\emm}_{S} = \Big|2\pi \sum_{S'\nu} \delta(\Omega_S-\Omega_{S'}-\omega_{\nu})\ \big|G_{S'S\nu}\big|^2 \Big|\ ,
\end{equation}
where we have introduced the exciton-phonon matrix element as $G_{S'S\nu}=\braket{S'|g_{\nu}|S}$, with $g_\nu = \braket{n_\nu+1|V_{ep}|n_\nu}$. This is the expected FGR expression for scattering between exciton states of the same reference Hamiltonian via interaction with phonons. The scattering time is therefore, $\tau_S = \gamma^{-1}_S$. The inverse lifetime derived here from the Lippmann-Schwinger scattering formalism is identical to twice the imaginary part of the first-order Fan-Migdal exciton-exciton-phonon self-energy (denoted FMd) for phonon emission~\cite{antonius2022theory}, namely
\begin{gather}
\begin{split}~\label{eq:se_FMd}
        \Xi^{\FMd}_{SS}(\Omega_S) &= \sum_{S'\nu} \frac{|G_{S'S\nu}|^2}{\Omega_S-\Omega_{S'}-\omega_{\nu}+i\eta} \ .
\end{split}
\end{gather}
Reinserting the wavevector dependence of all quantities and focusing only on zero momentum excitons, whose crystal momentum we denote by $\Gamma$, leads to the expression for the exciton emission scattering rate (xxFMd) as
\begin{equation}~\label{eq:x-x}
     \gamma^{\emm}_{\x\x\FMd} = \Big|2\pi \sum_{S'\mathbf{q}\nu} \delta(\Omega_S-\Omega_{S'\mathbf{q}}-\omega_{\mathbf{q}\nu})\ \big|G_{S'S\nu}(\Gamma,\mathbf{q})\big|^2 \Big|\ .
\end{equation}
One may perform a similar analysis for the absorption scattering pathway whereby an exciton absorbs a phonon and is scattered into another exciton state, yielding 
\begin{equation}
    \gamma^{\abs}_{\x\x\FMd} = \Big|2\pi \sum_{S'\mathbf{q}\nu} \delta(\Omega_S-\Omega_{S'\mathbf{q}}+\omega_{\mathbf{q}\nu})\ \big|G_{S'S\nu}(\Gamma,\mathbf{q})\big|^2 \Big|. 
\end{equation}
Including the temperature dependence introduces the relevant Bose-Einstein occupation factors and results in an expression identical with that derived from the finite temperature Matsubara formalism.~\cite{antonius2022theory,chen2020exciton,cudazzo2020first} 

%As stated in Ref~\cite{antonius2022theory}, the basis of valence and conduction bands used to expand the exciton wavefunction in equation~\ref{eq:tda}, in any practical implementation, will be incomplete and the computation of all possible exciton states becomes rapidly intractable. Hence, their work only considers the scattering between the first few exciton states under the assumption that the phonon frequency will be much lower than the exciton binding energy. However, in many materials, this may not be the case and the scattering by phonons to higher exciton states must be taken into account. 

%\subsection{Phonon-driven free electron-hole pair to free electron-hole pair scattering}\label{sub:eh-eh}

We may perform the same analysis for the free electron-hole to free electron-hole scattering channel ($V_{eh}=0$) by partitioning the Hamiltonian into a free electron-hole-phonon reference and an electron-phonon perturbation. In this case, the eigenstates of the reference Hamiltonian are now the free electron-hole-pair-phonon product states $\ket{vc,n_\nu}$, with eigenvalues $E^{vc,\nu} = (\epsilon_c-\epsilon_v)+n_\nu\omega_\nu$. Focusing on the single-phonon emission channel and following the same steps as defined previously for scattering between bound exciton states (see Appendix~\ref{app:ex_ex}), we can directly write the FGR inverse lifetime as 
\begin{equation}~\label{eq:eh-eh}
    \gamma^{\emm}_{vc} = \Big|2\pi\sum_{v'c',\nu} \delta[(\epsilon_c-\epsilon_v)-(\epsilon_{c'}-\epsilon_{v'})-\omega_{\nu}] \big|G_{v'c',vc,\nu}\big|^2 \Big| ,
\end{equation}
where we have taken the first Born approximation and defined the electron-hole-phonon vertex as $G_{v'c',vc,\nu}=\braket{v'c'|g_{\nu}|vc}$. The electron-hole-phonon vertex can be expanded in terms of the electron and hole phonon vertices as $G_{v'c',vc,\nu}=g_{c'c\nu}\delta_{v'v}-g_{vv'\nu}\delta_{c'c}$.~\cite{ring2004nuclear} Inserting this relation into Eq.~\ref{eq:eh-eh} while rotating into the exciton basis~\cite{marini2008ab} and reintroducing the wavevector dependence of all quantities, yields 
\begin{gather}
       \begin{split}\label{eq:approx_2_k}
        \gamma^{\emm}_{\UE} = \ 2\pi \sum_{vcc_1\nu\mathbf{k}\mathbf{q}} |A^{S}_{vc\mathbf{k}}g_{cc_1\nu}(\mathbf{k,\mathbf{q}})|^2\delta(\epsilon_{c\mathbf{k}}- \epsilon_{c_1\mathbf{k+q}}-\omega_{\mathbf{q}\nu})\\
        +2\pi \sum_{vcv_1\nu\mathbf{k}\mathbf{q}} |A^{S}_{vc\mathbf{k}}g_{v_1v\nu}(\mathbf{k},\mathbf{q})|^2\delta(\epsilon_{v_1\mathbf{k+q}}-\epsilon_{v\mathbf{k}}-\omega_{\mathbf{q}\nu})\ ,
    \end{split}
\end{gather}
where we have neglected the cross terms, which is expected to be an excellent approximation. Interestingly, as noted in Ref~\citenum{antonius2022theory}, this expression yields a non-zero lifetime of the exciton state at zero temperature, the consequences of which are explored in Section~\ref{sec:results}. We shall refer to this lifetime expression as the uncorrelated exciton (UE) approximation, in line with the nomenclature of Ref~\citenum{antonius2022theory}. 

The above expression, derived separately in Refs~\citenum{marini2008ab} and~\citenum{antonius2022theory}, is equivalent to twice the imaginary part of the diagonal independent electron-hole-pair-phonon self-energy (IEHPP) evaluated at the non-interacting electron-hole pair energy, $\epsilon_{c\mathbf{k}}-\epsilon_{v\mathbf{k}}$.~\cite{antonius2022theory} The uncorrelated exciton-self-energy is written as~\cite{antonius2022theory} 
\begin{equation}~\label{eq:ue_approx}
    \Xi^{\UE}_{SS}(\epsilon_c-\epsilon_v) = \sum_{vc}|A^{S}_{vc}|^2\big[\Sigma^{\FM}_{cc}(\epsilon_c)-\Sigma^{\FM}_{vv}(\epsilon_v)\big],
\end{equation}
where $\Sigma^{\FM}_{cc}(\omega)$ and $\Sigma^{\FM}_{vv}(\omega)$ are the usual Fan-Migdal electron-phonon self-energies for the conduction and valence bands.~\cite{giustino2017electron,antonius2022theory} An analogous expression can also be derived for the phonon absorption pathway by modifying the delta functions above and will be discussed in Section~\ref{sec:results}. Eq.~\ref{eq:approx_2_k} results from a picture of the exciton scattering rate originating solely from the broadening of the underlying single-particle states via the electron-phonon interaction.~\cite{marini2008ab,antonius2022theory} 

%The expressions derived (\ref{eq:x-x}) and (\ref{eq:approx_2_k}) are so far are based on the usual application of the Lippmann-Schwinger equations to generate FGR expressions for the exciton--exciton and free electron-hole -- electron-hole scattering pathways mediated by the electron-phonon interaction. In the next Sections we will explore a modified form of Fermi's Golden rule to account for scattering between states which are no longer of the same reference Hamiltonian. We will show that these processes are well defined within the theory of rearrangement collisions, allowing for the description of exciton scattering to free charge carriers. 

\section{Rearrangement Collision Theory for exciton Dissociation in Solids} \label{sec:rearrange}
Below, we first outline the partitionings of the Hamiltonian in Eq.~\ref{eq:ham}, and then we introduce our modified initial and final states, using rearrangement collision theory, to derive exciton to free charge carrier dissociation rates. 

\subsection{Partitioning the Exciton--Phonon Hamiltonian}~\label{sec:prior-post}
We start with the same partitioning as in Eq.~\ref{eq:initial}, and the initial scattering-state wavefunction is given by $\ket{\psi^+_{1i}} = \ket{S,n_\nu} + G^+(E^{S,\nu}_i)V_{ep}\ket{S,n_\nu}$, with scattering potential, $V_{ep}$. We want to describe the process of exciton scattering to a final state corresponding to a free electron-hole pair. The final state is now an eigenstate of a different reference Hamiltonian corresponding to the following partitioning 
\begin{subequations}
    \begin{equation}
        H^{(0)}_f = H_{ph} + H^{\IP}_{0} ,
    \end{equation}
    with ostensibly
\begin{equation}
    V_{f} = V_{eh} + V_{ep} \ ,
\end{equation}
\end{subequations}
mediating the scattering. Here, the electron-hole interaction enters via the final scattering operator, $V_f$. The Lippmann-Schwinger scattered final state is given by 
\begin{equation}\label{eq:final_eh}
    \ket{\psi^{-}_{2f}} = \ket{vc,n_\nu+1} + G^{-}(E^{vc,\nu}_{f})V_f\ket{vc,n_\nu+1} .
\end{equation}
The extended $\mathcal{S}$-matrix for a rearrangement collision process is defined in the usual way, but allowing for the different scattering channels,~\cite{sunakawa1960theory} $    \mathbf{S}_{2f,1i} = \braket{\psi^{-}_{2f}|\psi^+_{1i}}$, where the indices $2/1$ indicate the different Hamiltonian partitionings defining the different scattering channels. In regular $\mathcal{S}$-matrix theory, the $\mathcal{S}$-matrix is defined as in Eq.~\ref{eq:norm_s} with a single interaction common to both scattering channels. This means that no ambiguity arises when expanding the $\mathcal{S}$-matrix and taking the Born approximation. However, the extended $\mathcal{S}$-matrix must also obey the constraint of unitarity, as it represents the probability amplitude for a collision occurring in channel $1$ to be produced in channel $2$. A naive expansion of a general matrix element of the extended $\mathcal{S}$-matrix according to the conventional $\mathcal{S}$-matrix theory would give~\cite{sunakawa1960theory,sachs1987physics}
\begin{gather}
    \begin{split}
        \mathbf{S}_{2f,1i} &= -2\pi i \delta(E^{vc,\nu}_{f}-E^{S,\nu}_{i})\braket{vc,n_\nu'|V_{ep}+V_{eh}|\psi^{+}_{1i}}\\
        &= -2\pi i \delta(E^{vc,\nu}_{f}-E^{S,\nu}_{i})\braket{\psi^{+}_{2f}|V_{ep}| S,n_\nu} ,
    \end{split}
\end{gather}
where the first equality stems from evolving the initial state and projecting onto the final state while the second equality is obtained by evolving the final state and then projecting onto the initial state. This definition of the $\mathcal{S}$-matrix is neither unitary nor unique, and upon application of the Born approximation would give rise to the following ambiguity as
\begin{equation}
    \braket{vc,n_\nu'|V_{ep}+V_{eh}|S,n_\nu} \neq \braket{vc,n_\nu'|V_{ep}|S,n_\nu} \ .
\end{equation}
This is known as the `prior-post' discrepancy issue~\cite{lippmann1956rearrangement,epstein1957theory,sunakawa1960theory, day1961note,mittleman1961formal,chen1966formal,hahn1967theory,sachs1987physics} and arises due to the fact that the $\mathcal{S}$-matrix for this process is not well defined. We address this problem by extending the FGR treatment to rearrangement collisions.

%Therefore, special care has to be taken to define a unique, unitary extended $\mathcal{S}$-matrix that gives rise to the correct Born approximation, thereby avoiding the `prior-post' discrepancy issue. Alternatively, the naive expansion of the extended $\mathcal{S}$-matrix theory for the `rearrangement collision' of exciton dissociation to free electron-hole pairs may also be written as 
%\begin{equation}
%    \mathbf{S}_{2f,1i} = \braket{vc,n_\nu+1|U(V_f;\infty,0)U(V_{ep};0,-\infty)|S,n_\nu} ,
%\end{equation}
%where the time evolution operator is defined in the interaction picture as $U(V;t_2,t_1) = \mathcal{T} \left[\exp\left(-i\int^{t_2}_{t_1}dt\ V(t)\right)\right]$.~\cite{gell1951bound} In this framework, it is not possible to define a single scattering operator which is common to both scattering channels. 

%In what follows, we introduce a modification of the initial and/or final states such that we can define a common scattering operator to every scattering channel, allowing us to take the correct Born approximation, there

\subsection{Modifying initial and final states and the generalized optical theorem}
From Eqs~\ref{eq:lipp} and \ref{eq:final_eh}, we see that the common scattering operator between all channels is given by the electron-phonon interaction $V_{ep}$, whilst the electron-hole interaction $V_{eh}$, prevents us from defining a common scattering operator connecting initial and final states. Following the standard rearrangement collision theory strategy (as shown in Ref.~\citenum{sunakawa1960theory}), we modify the free electron-hole state such that it is defined with respect to the same initial partitioning and scattering operator.

From Eq.~\ref{eq:initial}, we may write the Green's function as a Dyson series~\cite{sunakawa1960theory,mittleman1961formal,epstein1957theory} 
\begin{equation}~\label{eq:dys}
    G^{\pm}(E) = G^{\pm}_0(E) + G^\pm(E)V_{ep}G^{\pm}_0(E) \ ,
\end{equation}
where $G^{\pm}_0(E) = \left(E-H_0\pm i\eta\right)^{-1}$ is the reference Green's function corresponding to the reference Hamiltonian in Eq.~\ref{eq:initial}. Using Eq.~\ref{eq:dys}, we write the final state of Eq.~\ref{eq:final_eh} as
\begin{equation}
    \ket{\psi^-_{2f}} = \ket{\chi^-_{2f}} + G^-(E^{vc,\nu}_{f})V_{ep}\ket{\chi^-_{2f}} \ ,
\end{equation}
where we have defined the renormalized electron-hole state as (see Appendix~\ref{app:re} for details)
\begin{equation}
    \ket{\chi^-_{2f}} = \ket{vc,n_\nu+1} + G^{-}_0(E^{vc,\nu}_f)V_{eh}\ket{vc,n_\nu+1} \ .
\end{equation}
The electron-hole interaction has thus been absorbed into the renormalized state $\ket{\chi^-_{2f}}$, which is now an eigenstate of the same initial reference Hamiltonian, $H_0$. One may think of this state as an orthogonalized free electron-hole pair (orthogonalized with respect to the bound exciton states), a more rigorous alternative to the orthogonal plane waves introduced in Ref.~\citenum{perea2021phonon} to describe the continuum states, since our procedure results from an exact treatment of the quantum mechanical scattering. With this modified state, we may express the final scattered state to be evolved from the common $V_{ep}$ interaction as~\cite{sunakawa1960theory} 
\begin{equation}
    \ket{\psi^-_{2f}} = U(V_{ep};0,\infty)\ket{\chi^-_{2f}} \ .
\end{equation}
This relation allows us to write the expression for the generalized $\mathcal{S}$-matrix above as 
\begin{gather}
    \begin{split}
        \mathbf{S}_{2f,1i} &=  \braket{\chi^-_{2f}|U(V_{ep};\infty,0)U(V_{ep};0,-\infty)|S,n_\nu}\\
        &= \braket{\chi^-_{2f}|\mathcal{S}(V_{ep})|S,n_\nu} \ ,
    \end{split}
\end{gather}
where we have used the relation $\mathcal{S}(V_{ep}) = U(V_{ep};\infty,0)U(V_{ep};0,-\infty)$. 

Now that we have a well defined $\mathcal{S}$-matrix in terms of the electron-phonon interaction, we may simplify the expression further to give (see Appendix~\ref{app:gen} for details)
\begin{gather}
    \begin{split}~\label{eq:conser}
        \mathbf{S}_{2f,1i} &= -2\pi i \ \delta(E^{vc,\nu}_{f}-E^{S,\nu}_{i}) \braket{\chi^-_{2f}|V_{ep}|\psi^+_{1i}}  \ .
    \end{split}
\end{gather}
From the identity that the rate of transition from the initial exciton state to the final renormalized electron-hole pair, $w_{2f,1i}$, is given by~\cite{lippmann1950variational,hahn1967theory,chen1966formal} 
\begin{equation}~\label{eq:rate_1}
    w_{2f,1i} = \frac{\partial}{\partial t} \Big|\braket{\chi^{-}_{2f}|U(V_{ep};t,-\infty)|S,n_\nu}\Big|^2 \ ,
\end{equation}
it may be shown~\cite{sunakawa1960theory,lippmann1950variational} that this rate reduces to (Appendix~\ref{app2})
\begin{equation}\label{eq:general}
 \gamma_{S} = 2\pi\Big|\sum_{vc\nu}\delta(E^{vc,\nu}_{f}-E^{S,\nu}_{i})\ |\braket{\chi^-_{2f}|V_{ep}|\psi^+_{1i}}|^2\Big| \ ,
\end{equation}
where we have identified the exciton dissociation rate as equal to sum over all final states of the rate of transition from an exciton state to the final free electron-hole pair states, $\big|\sum_{2f}w_{2f,1i}\big|\equiv\gamma_{S}$.~\cite{mahan2000many,Peskin1995}

\subsection{Phonon-driven exciton dissociation}~\label{sub:diss}
Eq.~\ref{eq:general} contains the exact scattered states $\ket{\psi^+_{1i}}$ and $\ket{\chi^{-}_{2f}}$ which enter via the inner product $\braket{\chi^-_{2f}|V_{ep}|\psi^+_{1i}}$. To generate an approximate expression to first-order for the scattering rate, we employ the first Born approximation for this matrix element, which may be expanded using the modified Lippmann-Schwinger equation for the final state: 
\begin{gather}
    \begin{split}~\label{eq:modifi}
        \braket{\chi^-_{2f}|V_{ep}|\psi^+_{1i}} &= \braket{vc,n_\nu+1|V_{ep}|\psi^+_{1i}} \\
        &+ \braket{vc,n_{\nu}+1|V_{eh}G^{+}_0(E^{vc,\nu}_f)V_{ep}|\psi^+_{1i}} \ ,
    \end{split}
\end{gather}
where we have used the definition $\bra{\chi^{-}_{2f}} = \bra{vc,n_\nu+1}+\bra{vc,n_\nu+1}V_{eh}G^{+}_0(E^{vc,\nu}_f)$. From the energy conservation condition above (Eq.~\ref{eq:conser}), we replace $E^{vc,\nu}_f$ with $E^{S,\nu}_i$, and using the definition of $\ket{\psi^+_{1i}}$, we rearrange Eq.~\ref{eq:modifi} as
\begin{gather}
    \begin{split}
        \braket{\chi^-_{2f}|V_{ep}|\psi^+_{1i}} &= \braket{vc,n_\nu+1|V_{ep}+V_{eh}|\psi^+_{1i}} \\
        &-\braket{vc,n_\nu+1|V_{eh}|S,n_\nu} \ .
    \end{split}
\end{gather}
Taking the first Born approximation, $\ket{\psi^+_{1i}}\approx\ket{S,n_\nu}$, we get 
\begin{equation}
    \braket{\chi^-_{2f}|V_{ep}|\psi^+_{1i}} \approx \braket{vc,n_\nu+1|V_{ep}|S,n_\nu} \ .
\end{equation}
Remarkably, here we see that the correct first Born approximate expression for the scattering rate only includes the electron-phonon interaction. Identifying the electron-phonon coupling vertex, $g_\nu$ (as in Subsection~\ref{sub:ex-ex}), the initial and final energies, we find the expression for the rate (within the Born approximation), namely 
\begin{equation}
    \gamma^{\emm}_{S} = 2\pi\Big|\sum_{vc\nu}\delta(\Omega_S-(\epsilon_c-\epsilon_v)-\omega_\nu)\ |\braket{vc|g_{\nu}|S}|^2\Big| \ .
\end{equation}
This expression is the correct Born approximation for the exciton dissociation rate and here we have fully justified its use, avoiding the `prior-post' discrepancy issue outlined in Section~\ref{sec:prior-post}. The scattering matrix element within the Born approximation contains different contributions, which can be seen by resolving the identity of exciton states, $\braket{vc|g_{\nu}|S} = \sum_{v_2c_2} A^{S}_{v_2c_2} \left[g_{cc_2\nu}\delta_{vv_2}-g_{v_2v\nu}\delta_{cc_2}\right]$,
and expanding the square modulus:
\begin{widetext}
    \begin{gather}
    \begin{split}\label{eq:expanded}
        \gamma^{\emm}_S = & \Big|\ 2\pi \sum_{vc,c'c_1\nu} A^{S*}_{vc}g_{cc_1\nu}g^*_{c'c_1\nu}A^S_{vc'}\delta(\Omega_S-(\epsilon_{c_1}-\epsilon_v)-\omega_{\nu}) +2\pi \sum_{vc,v'v_1\nu} A^{S*}_{vc}g_{v_1v\nu}g^*_{v_1v'\nu}A^S_{v'c}\delta(\Omega_S-(\epsilon_{c}-\epsilon_{v_1})-\omega_{\nu})\\
        &-2\pi\sum_{vc,v'c',\nu} A^{S*}_{vc}g_{cc'\nu}g^*_{vv'\nu}A^{S}_{v'c'} \Big[\delta(\Omega_S-(\epsilon_c-\epsilon_{v'})-\omega_{\nu})+\delta(\Omega_S-(\epsilon_{c'}-\epsilon_v)-\omega_{\nu})\Big] \Big| \ .
    \end{split}
\end{gather}
\end{widetext}

\begin{figure}[ht]
	\includegraphics[width=85mm, height = 35mm]{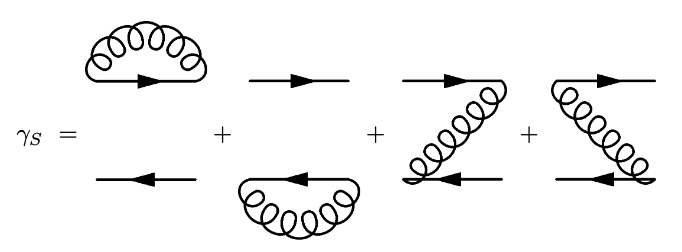} 
	\caption[]{Diagrams for the exciton dissociation rate within the first Born approximation. The first two diagrams correspond to the Fan-Migdal self-energy, while the last two diagrams are referred to as phonon exchange diagrams.}
	\label{fig:Kph}
\end{figure}
These terms can be represented by the Feynman diagrams displayed in Figure~\ref{fig:Kph}. Similar steps can be applied for the absorption channel to obtain, $\gamma^{\abs}_{S} = 2\pi\big|\sum_{vc\nu}\delta(\Omega_S-(\epsilon_c-\epsilon_v)+   \omega_\nu)\ |\braket{vc|g_{\nu}|S}|^2\big|$.

%From expression \ref{eq:expanded}, we can clearly see the relationship to the first-order MBPT self-energy. The terms appearing are similar to those which are obtained from the imaginary part of the independent electron-hole-pair-phonon self-energy (IEHPP) in the exciton basis~\cite{antonius2022theory}. However, there is a subtle difference due to the fact that this kernel is actually evaluated at the exciton eigenvalue, $\Omega_S$, not at the energy of the free electron-hole pair, $(\epsilon_c-\epsilon_v)$. From the arguments presented here, it is clear that the imaginary part of the IEHPP self-energy evaluated at the exciton energy describes the dissociation of excitons to free electron-hole pairs
%\begin{equation}
%	\gamma_{S} = 2\Big| \I \Xi^{\IEHPP}_{SS}(\Omega_S)\Big| \ .
%\end{equation}
\noindent The first two terms of Eq.~\ref{eq:expanded} originate from the Fan-Migdal (FM) electron-phonon self-energy expressed in the exciton basis. Within the weak coupling regime, we refer to electron- and hole-polarons as the quasiparticle solutions of the FM self-energy.~\cite{mahan2000many} The imaginary parts of the electron-phonon and hole-phonon FM self-energy in the free electron-hole basis (at zero temperature) are given by~\cite{giustino2017electron,antonius2022theory,mahan2000many}
\begin{subequations}
    \begin{equation}
        \I \Sigma^{\FM}_{cc'}(\omega)\delta_{vv'} = \pi\sum_{c_1\nu} g_{cc_1\nu}g^*_{c'c_1\nu} \delta_{vv'}\delta(\omega-\epsilon_{c_1}-\omega_\nu) \ ,
    \end{equation} 
and 
\begin{equation}
    \I \Sigma^{\FM}_{vv'}(\omega)\delta_{cc'} = \pi\sum_{v_1\nu} g_{v'v_1\nu}g^*_{vv_1\nu} \delta_{cc'}\delta(\omega-\epsilon_{v_1}+\omega_\nu) \ .
\end{equation}
\end{subequations}
Transforming these quantities to the exciton basis and taking their linear combination, we can relate the lifetime of the electron- and hole-polaron states due to the electron-phonon and hole-phonon interactions (for the emission channel) as
\begin{gather}
	\begin{split}
		\gamma^{\emm}_{\eh\FMd} &= 2\Big|\sum_{vc,v'c'}A^{S*}_{vc}\I \Sigma^{\FM}_{cc'}(\Omega_S+\epsilon_v)\delta_{vv'} A^S_{v'c'}\\
		&+\sum_{vc,v'c'}A^{S*}_{vc} \I \Sigma^{\FM}_{vv'}(-(\Omega_S-\epsilon_c))\delta_{cc'} A^S_{v'c'}\Big| \ .
	\end{split}
\end{gather}
We have introduced the notation ehFMd to indicate this dissociation pathway. Therefore, the expression for the inverse lifetime for exciton scattering to free electron-hole pairs becomes 
   \begin{gather}
	\begin{split}\label{eq:relabel}
		&\gamma^{\emm}_{S} = \Big|\gamma^{\emm}_{\eh\FMd}
		-2\pi\sum_{vc,v'c'\nu} A^{S*}_{vc}g_{cc'\nu}g^*_{vv'\nu}A^{S}_{v'c'}\\
		&\Big[\delta(\Omega_S-(\epsilon_c-\epsilon_{v'})-\omega_{\nu})+\delta(\Omega_S-(\epsilon_{c'}-\epsilon_v)-\omega_{\nu})\Big] \Big| \ .
	\end{split}
\end{gather}
From this expression we may identify two approximate regimes. First, if we assume that the polaron renormalization introduced by the Fan-Migdal self-energy results in (infinitely) long lived electron/hole-polaron quasiparticle states, we may simplify the exciton-phonon inverse lifetime in Eq.~\ref{eq:relabel} by neglecting the imaginary part of the $\FM$ self-energy. This would give the expression
\begin{gather}
	\begin{split}\label{eq:approx}
		&\gamma^{\emm}_{\x\eh} = 2\pi\Big|\sum_{vc,v'c',\nu} A^{S*}_{vc}g_{cc'\nu}g^*_{vv'\nu}A^{S}_{v'c'}\\
		&\Big[\delta(\Omega_S-(\epsilon_c-\epsilon_{v'})-\omega_{\nu})+\delta(\Omega_S-(\epsilon_{c'}-\epsilon_v)-\omega_{\nu})\Big]\Big| \ ,
	\end{split}
\end{gather}
where the notation xeh indicates the explicit rate of exciton dissociation to free electron-hole pairs. In this picture, the phonon exchange timescale is much faster than the individual electron and hole scattering lifetimes. Second, we may also make the approximation in which we neglect the phonon exchange term such that the correlated motion of the electron and hole is not influenced by the phonon field. This corresponds to a picture where the electron and hole scattering lifetimes are much faster than the phonon exchange timescale. This results in a modified form of the lifetime expression for the independent electron-hole pair from Subsection~\ref{sub:ex-ex}
\begin{gather}
	\begin{split}\label{eq:novel}
		\gamma^{\emm}_{\eh\FMd} &= 2\Big|\sum_{vcc'}A^{S*}_{vc} \I \Sigma^{\FM}_{cc'}(\Omega_S+\epsilon_v) A^S_{vc'}\\
		&+\sum_{vcv'}A^{S*}_{vc}\I \Sigma^{\FM}_{vv'}(-(\Omega_S-\epsilon_c))A^S_{v'c} \Big| \ .
	\end{split}
\end{gather}
Furthermore, by taking a diagonal approximation, we can write the scattering rate as
\begin{gather}
	\begin{split}\label{eq:approx_2}
		\gamma^{\emm}_{\eh\FMd} = &\Big|\ 2\pi \sum_{vcc_1\nu} |A^{S}_{vc}g_{cc_1\nu}|^2\delta(\Omega_S-(\epsilon_{c_1}-\epsilon_v)-\omega_{\nu})\\
		+&2\pi \sum_{vcv_1\nu} |A^{S}_{vc}g_{v_1v\nu}|^2\delta(\Omega_S-(\epsilon_{c}-\epsilon_{v_1})-\omega_{\nu})\Big| \ , 
	\end{split}
\end{gather}
a modified form of the lifetime obtained from the IEHPP self-energy evaluated at the exciton energy as opposed to the independent electron-hole pair energy. The self-energy corresponding to the uncorrelated exciton dissociation rate may be explicitly written as 
\begin{equation}~\label{eq:ehFMd_se}
    \Xi^{\UE}_{SS}(\Omega_S) = \sum_{vc}|A^{S}_{vc}|^2\big[\Sigma^{\FM}_{cc}(\Omega_S+\epsilon_v)-\Sigma^{\FM}_{vv}(-(\Omega_S-\epsilon_c))\big]\ .
\end{equation}
In the non-interacting limit where $\Omega_S\approx\epsilon_c-\epsilon_v$, Eq.~\ref{eq:ehFMd_se} reduces to the UE approximaton of Eq.~\ref{eq:ue_approx}. In Subsection~\ref{sub:ex-pol} we will show how Eq.~\ref{eq:approx} naturally emerges from a different partitioning of the Hamiltonian and represents the explicit exciton dissociation lifetime.  

To the best of our knowledge, Eqs~\ref{eq:approx} and~\ref{eq:approx_2} have not been reported thus far. In particular, Eq.~\ref{eq:approx_2} describes the scenario in which phonons strongly alter the character of the excited electron or hole states. Previous work by Strinati on the broadening of excitons due to the electron-hole Coulomb interaction identified similar contributions resulting from either the exciton or underlying band structure.~\cite{strinati1982dynamical} The scattering rates derived thus far are summarized in Table~\ref{table:expressions}.

\subsection{Exciton-polaron to free electron-hole-polaron pair scattering}\label{sub:ex-pol}

Following the derivation for the dissociation rate of excitons to free electron-hole pairs, it is possible to extend this analysis further to exciton-polaron dissociation into free electron-hole-polaron pairs. In this case, the reference Hamiltonians are all Hermitian, with real eigenvalues and orthogonal eigenstates. In the description of exciton-polarons, the effect of the electron-phonon interaction is explicitly included in the independent electron-hole pair Hamiltonian, $H^{\IP}_0$. This corresponds to the picture of polaronic band renormalization,~\cite{giustino2017electron,lafuente2022ab} which can be seen by writing the electron-phonon interaction in the free electron-hole pair basis, $\Tilde{V}_{ep}$, as
\begin{equation}
	\Tilde{V}_{ep} = V_{elp} + V_{hp} + V_{elhp}
\end{equation}
where $V_{elp}$ represents the `electron'-phonon interaction, $V_{hp}$, the `hole'-phonon interaction and $V_{elhp}$, the combined `electron-hole'-phonon interaction. Here, one may think of the electron-phonon and hole-phonon interactions as resulting in the polaronic renormalization of the electronic bands~\cite{giustino2017electron,lafuente2022ab,lafuente2022unified} that is now absorbed by $H^{\IP}_0$. The combined electron-hole-phonon interaction represents the coupled electron and hole motion due to the phonon field. Therefore, we define the initial exciton state to be constructed from the renormalized polaron bands (forming an exciton-polaron) and write the initial partitioning as 
\begin{subequations}
    \begin{equation}
        H_0 = H_{ph} + \Tilde{H}^\IP_{0}+ V_{eh}\ ,
    \end{equation}
with the scattering potential 
\begin{equation}
    V_0 = V_{elhp} \ . 
\end{equation}
\end{subequations}
\begin{table*}[ht]
	\setlength{\tabcolsep}{6pt} % Default is 6pt
	\begin{tabular}{llll} % Was {lcr}
        \hline
		\hline \\[-5pt]
  		Scattering  & \emph{Ab Initio} Scattering Rate (phonon absorption) & \vtop{\hbox{\strut Self-Energy} \hbox{\strut Kernel}} & Physical Interpretation \\[5pt]
  		\hline\\[-5pt]  
		\vtop{\hbox{\strut xxFMd}\hbox{\strut (Eq.~\ref{eq:x-x})}}  & $2\pi\sum_{S'\nu}|G_{S'S\nu}|^2 N_B(\omega_\nu)\delta(\Omega_S-\Omega_{S'}+\omega_{\nu})$  \vspace{1mm}  & \vtop{\hbox{\strut $\Xi^{\FMd}_{SS}(\Omega_S)$}\hbox{\strut (Eq.~\ref{eq:se_FMd})}} & \vtop{\hbox{\strut Scattering between exciton}\hbox{\strut states} }\\[5pt]
		\vtop{\hbox{\strut xeh}\hbox{\strut (Eq.~\ref{eq:pol}) } } & \vtop{\hbox{\strut $2\pi\sum_{\substack{vc\\v'c'\nu}}A^{S*}_{vc}g_{cc'\nu}g^*_{vv'\nu}A^{S}_{v'c'}N_B(\omega_\nu)\delta(\Omega_S-(\epsilon_c-\epsilon_{v'})+\omega_\nu)+ $}\hbox{\strut $2\pi\sum_{\substack{vc\\v'c'\nu}}A^{S*}_{vc}g_{cc'\nu}g^*_{vv'\nu}A^{S}_{v'c'}N_B(\omega_\nu)\delta(\Omega_S-(\epsilon_{c'}-\epsilon_{v})+\omega_\nu)$}}  \vspace{1mm} & \vtop{\hbox{\strut $K^{ph}_{SS}(\Omega_S)$}\hbox{\strut (Eq.~\ref{eq:kph})}} & \vtop{\hbox{\strut Dissociation of a  bound }\hbox{\strut exciton to continuum }\hbox{\strut electron-hole states}} \\[5pt]
		\vtop{\hbox{\strut ehFMd } \hbox{\strut(Eq.~\ref{eq:approx_2})}} & \vtop{\hbox{\strut $2\pi \sum_{vcc'\nu} |A^{S}_{vc}g_{cc'\nu}|^2N_B(\omega_\nu)\delta(\Omega_S-(\epsilon_{c'}-\epsilon_v)+\omega_{\nu}) + $}\hbox{\strut $2\pi \sum_{vcv'\nu} |A^{S}_{vc}g_{v'v\nu}|^2N_B(\omega_\nu)\delta(\Omega_S-(\epsilon_{c}-\epsilon_{v'})+\omega_{\nu})$}}   \vspace{1mm}   & \vtop{\hbox{\strut $\Xi^{\UE}_{SS}(\Omega_S)$}\hbox{\strut (Eq.~\ref{eq:ehFMd_se})}} &\vtop{\hbox{\strut Uncorrelated dissociation to }\hbox{\strut continuum electron-hole}\hbox{\strut pairs}} \\[5pt]
        \vtop{\hbox{\strut UE}\hbox{\strut(Eq.~\ref{eq:approx_2_k})}} & \vtop{\hbox{\strut $2\pi \sum_{vcc'\nu} |A^{S}_{vc}g_{cc'\nu}|^2 N_B(\omega_\nu)\delta(\epsilon_{c}-\epsilon_{c'}+\omega_{\nu}) +$}\hbox{\strut $2\pi \sum_{vcv'\nu}|A^{S}_{vc}g_{vv'\nu}|^2 N_B(\omega_\nu)\delta(\epsilon_{v'}-\epsilon_{v}+\omega_{\nu}) $}} \vspace{1mm} & \vtop{\hbox{\strut$\Xi^{\UE}_{SS}(\epsilon_c-\epsilon_v)$}\hbox{\strut (Eq.~\ref{eq:ue_approx})}} &\vtop{\hbox{\strut Scattering between }\hbox{\strut uncorrelated electron-hole}\hbox{\strut pairs}}
        \\[5pt]
        \hline
        \hline
	\end{tabular}
	\caption{Summary of the different scattering channels presented in this work, their associated exciton-phonon self-energy kernels and physical interpretation. xxFMd indicates the exciton--exciton scattering rate; xeh stands for the explicit exciton dissociation rate; ehFMd is the dissociation rate due to the broadening of the uncorrelated electron and hole states; and UE is the uncorrelated exciton approximation where the exciton is treated as an independent electron-hole pair.}
	\label{table:expressions}
\end{table*}
Here, $\Tilde{H}^{\IP}_0 = H^\IP_0 + V_{elp} + V_{hp}$ explicitly includes the effects of polaronic band renormalization on the single-particle electronic states. The exciton-polaron eigenstate and energy solutions are given by $\ket{\Tilde{S}}$ and $\Tilde{\Omega}_S$, respectively. Our final state will correspond to the free electron-hole-polaron state arising from the partitioning 
\begin{subequations}
    \begin{equation}
        H_f = H_{ph} + \Tilde{H}^\IP_{0}
    \end{equation}
and
\begin{equation}
    V_{f} = V_{elhp} + V_{eh} \ ,
\end{equation}
\end{subequations}
with eigenstate and energy solutions, $\ket{\Tilde{vc}}$ and $(\Tilde{\epsilon}_c-\Tilde{\epsilon}_v)$, respectively. Following exactly the same arguments as those outlined in Section~\ref{sub:diss}, the electron-hole interaction may be absorbed into the modified final state. The resulting first Born approximation for the rate of exciton-polaron dissociation to free electron-hole-polaron pairs via phonon absorption/emission is written as 
\begin{gather}
	\begin{split}~\label{eq:diss_pol}
		\gamma_{S} = 2\pi\Big|\sum_{vc\nu}\delta(\Tilde{\Omega}_S-(\Tilde{\epsilon}_c-\Tilde{\epsilon}_v)\pm\omega_\nu)\ |\braket{\Tilde{vc}|V^{elhp}_{\nu}|\Tilde{S}}|^2\Big| \ ,
	\end{split}
\end{gather}
where $V^{elhp}_\nu=\braket{n_\nu+1|V_{elhp}|n_\nu}$. Eq.~\ref{eq:diss_pol} contains the explicit phonon-driven dissociation rate of an exciton--polaron into its composite electron-hole-polaron pair. 

At this point, we can exactly recover the expression in Ref.~\citenum{alvertis} for the exciton dissociation lifetime by neglecting the effects of the polaronic renormalization of the band structure due to the electron-phonon and hole-phonon interactions. Depending on the nature of the interference between the electron and hole polarons, their renormalized effects may cancel each other out within this regime. As a first approximation, previous work partially neglects the effects of this band renormalization entirely.~\cite{mahanti1972effective,alvertis,filip2021phonon} Taking this approximation here, the expression for the scattering rate becomes 
\begin{gather}
	\begin{split}
		\gamma_{S} = 2\pi\Big|\sum_{vc\nu}\delta(\Omega_S-(\epsilon_c-\epsilon_v)\pm\omega_\nu)\ |\braket{vc|V^{elhp}_{\nu}|S}|^2\Big| \ ,
	\end{split}
\end{gather}
where the wavefunctions and energies are replaced by the exciton and free electron-hole counterparts. Expanding the electron-hole-phonon interaction, $V^{elhp}_{\nu}$, within the first Born approximation gives twice the imaginary part of the first-order phonon exchange term (named $K^{ph}_{SS}$ in Refs~\citenum{filip2021phonon} and~\citenum{alvertis}) in the exciton basis evaluated at the exciton energy, namely
\begin{gather}
	\begin{split}~\label{eq:kph}
		&K^{ph}_{SS}(\Omega_S) = -\sum_{\substack{vc,v'c'\nu\\}} A^{S*}_{vc}g_{cc'\nu}g^*_{vv'\nu}A^S_{v'c'}\\
  &\left[\frac{1}{\Omega_S-(\epsilon_c-\epsilon_{v'})-\omega_\nu+i\eta} + \frac{1}{\Omega_S-(\epsilon_{c'}-\epsilon_{v})-\omega_\nu+i\eta}\right] .
	\end{split}
\end{gather}
This is exactly the expression obtained in Section~\ref{sub:diss} by ignoring the effects of the Fan Migdal renormalization of the band structure (Eq.~\ref{eq:approx}), i.e. assuming single-particle states electron and hole states of infinite lifetime. Reinserting the wavevector dependence of all quantities, we obtain
\begin{widetext}
    \begin{gather}
	\begin{split}~\label{eq:pol}
		\gamma_{\x\eh} = 2\pi\Big| \sum_{  \substack{vc\mathbf{k}\\
v'c'\mathbf{k'}\nu}} A^{S*}_{vc\mathbf{k}}g_{cc'\nu}(\mathbf{k'},\mathbf{q})g^*_{vv'\nu}(\mathbf{k'},\mathbf{q})A^{S}_{v'c'\mathbf{k'}}\Big[\delta(\Omega_S-(\epsilon_{c\mathbf{k}}-\epsilon_{v'\mathbf{k'}})-\omega_{\mathbf{q}\nu})
		+\delta(\Omega_S-(\epsilon_{c'\mathbf{k'}}-\epsilon_{v\mathbf{k}})-\omega_{\mathbf{q}\nu})\Big]\Big| ,
	\end{split}
\end{gather}
\end{widetext}
where $\mathbf{k}=\mathbf{k'+q}$ due to conservation of crystal momentum. The expression for phonon absorption is trivially found by flipping the sign of the phonon frequency above. From this expression, we can identify the lifetime of exciton dissociation to free electron-hole pairs as due to the combined `electron-hole'-phonon interaction corresponding to the first-order exchange diagrams (Figure~\ref{fig:Kph}), which are responsible for the correlated motion of the electron and hole in the presence of the phonon field.~\cite{filip2021phonon,alvertis} This derivation  demonstrates that the dissociation rate of an exciton--polaron to a free electron-hole-polaron pair (the physical process that occurs in real materials) must result from the phonon exchange term where the single-particle broadenings are neglected. Therefore, Eq.~\ref{eq:pol} represents the \emph{explicit} phonon-driven exciton dissociation lifetime. 

One may include the temperature dependence of all scattering lifetimes derived via the Bose-Einstein occupation factors for the phonons, since excitation energies are much larger than the phonon frequencies.~\cite{antonius2022theory,alvertis,chan2023exciton} The emission channel acquires the term $N_{B}(\omega_{\mathbf{q}\nu})+1$, whilst the absorption term acquires $N_{B}(\omega_{\mathbf{q}\nu})$, where $N_{B}(\omega_{\mathbf{q}\nu})=(e^{\beta\omega_{\mathbf{q}\nu}}-1)^{-1}$ is the Bose-Einstein occupation factor and $\beta = \frac{1}{k_BT}$ is the inverse temperature. 

%We note in passing that the broad applicability of rearrangement collision theory also allows us to re-derive the optical--elemental exciton scattering lifetime recently reported in Ref.~\cite{paleari2022exciton} (see Appendix~\ref{app3}). 

\section{Application to a model system}\label{sec:results}

We demonstrate the applicability of our formalism by calculating the exciton dissociation rate for a model system based on the hydrogenic model for excitons and Fröhlich model for electron-phonon interactions in the same spirit as Ref.~\citenum{filip2021phonon}. This general model for exciton scattering rates is based on several characteristic material parameters such as the longitudinal optical (LO) phonon frequency, $\omega_{LO}$, and the exciton binding energy, $E_B$. 

\subsection{Hydrogenic excitons, Fröhlich interaction and model expressions}~\label{sub:models}
Excitons in many three-dimensional polar semi-conductors can be modelled accurately within the hydrogenic (Wannier-Mott) approximation,~\cite{alvertis,filip2021phonon} whereby the lowest energy (1s) state leads to the following analytic expression for the exciton coefficients $A_{\mathbf{k}} = \frac{(2a_0)^{\frac{3}{2}}}{\pi}\cdot\frac{1}{(1+a^2_0|\mathbf{k}|^2)^2}$, where $a_0 = \sqrt{\frac{1}{2E_B\mu}}$ is the exciton Bohr radius, with $E_B =E_g-\Omega_{1s}$, the exciton binding energy for the 1s state and $\mu=\frac{m_em_h}{m_e+m_h}$, the exciton reduced mass, and where $m_e$ is the conduction band effective mass and $m_h$ the valence band effective mass. 

The Fröhlich interaction is a result of the coupling between the LO phonon mode and the electronic degrees of freedom,~\cite{doi:10.1080/00018735400101213} with the expression for the real space operator given by $g^{F}_{\mathbf{q}}(\mathbf{r}) = \frac{i}{|\mathbf{q}|}\sqrt{\frac{2\pi \omega_{LO}}{NV}\left(\frac{1}{\epsilon_{\infty}}-\frac{1}{\epsilon_0}\right)}e^{i\mathbf{q\cdot r}}$,~\cite{doi:10.1080/00018735400101213,filip2021phonon, alvertis}  where $V$ is the unit cell volume, $N$ is the number of unit cells, $\epsilon_{\infty/0}$ are the high/low frequency dielectric constants and $\omega_{LO}$ is the frequency of the dispersionless LO phonon responsible for the Fröhlich interaction. 

Assuming parabolic conduction and valence bands and explicitly extending our results to finite temperature, we can write model expressions for the scattering rates of the 1s exciton state for the different scattering channels derived. The exciton dissociation rate (Eq.~\ref{eq:pol}) may be simplified by noting that the energy conservation enforced by the delta functions is satisfied only if $\omega_{LO}> E_B$ for our model system.~\cite{alvertis} Additionally, the exciton scattering rate at zero temperature for the 1s state will be exactly zero as the exciton cannot scatter into a lower energy state via phonon emission (unless via an indirect band gap pathway, not discussed here). The only non-zero contribution to the exciton dissociation rate will result from the phonon absorption scattering channel, namely 
\begin{gather}
	\begin{split}\label{eq:numerical}
		\gamma_{\x\eh}(T) = 2\pi \sum_{\mathbf{k q}} &A^*_{\mathbf{k+q}}|g^{F}_{\mathbf{q}}|^2 A_{\mathbf{k}}N_B(\omega_{LO})\\
		&\left[\delta\left( E_B+\frac{|\mathbf{k+q}|^2}{2m_e}+\frac{|\mathbf{k}|^2}{2m_h}-\omega_{LO}\right)  \right. \\
		& \left.  +\delta\left(E_B+\frac{|\mathbf{k}|^2}{2m_e}+\frac{|\mathbf{k+q}|^2}{2m_h}-\omega_{LO}\right) \right] \ ,
	\end{split}
\end{gather}
where $N_B(\omega_{LO}) = (e^{\beta\omega_{LO}}-1)^{-1}$ is the Bose-Einstein occupation factor for the LO mode.

The model expression for the approximate scattering channel where the phonon exchange term is neglected and exciton dissociation is based on a single-particle broadening as in Eq.~\ref{eq:approx_2} gives 
\begin{gather}
    \begin{split}\label{eq:FM}
        \gamma_{\eh\FMd}(T) = 2\pi \sum_{\mathbf{k q}} &|A_{\mathbf{k}}|^2|g^{F}_{\mathbf{q}}|^2 N_B(\omega_{LO})\\
        &\left[\delta\left( E_B+\frac{|\mathbf{k+q}|^2}{2m_e}+\frac{|\mathbf{k}|^2}{2m_h}-\omega_{LO}\right)  \right. \\
        & \left.  +\delta\left(E_B+\frac{|\mathbf{k}|^2}{2m_e}+\frac{|\mathbf{k+q}|^2}{2m_h}-\omega_{LO}\right) \right]  ,
    \end{split}
\end{gather}
The related uncorrelated exciton approximation, derived in Eq.~\ref{eq:approx_2_k}, takes the form
\begin{gather}
    \begin{split}
\gamma_{\UE}(T) = 2\pi \sum_{\mathbf{k q \pm}} |A_{\mathbf{k}}|^2|g^{F}_{\mathbf{q}}|^2 (N_B(\omega_{LO})+\frac{1}{2}\pm\frac{1}{2})\\
\left[\delta\left( \frac{|\mathbf{k}|^2}{2m_e}-\frac{|\mathbf{k+q}|^2}{2m_e}\mp\omega_{LO}\right) \right. \\
\left. + \delta\left( \frac{|\mathbf{k}|^2}{2m_h}-\frac{|\mathbf{k+q}|^2}{2m_h}\mp\omega_{LO}\right) \right] \ ,
    \end{split}
\end{gather}
where the $\pm$ notation indicates the sum over both emission and absorption channels. We include both absorption and emission channels as the UE approximation does not account for the exciton binding energy meaning that the energy conservation condition for the emission channel is also satisfied at all temperatures. As a result, this expression clearly implies a non-zero rate at zero temperature. 

%\begin{figure*}[ht]
%	\centering
%	\subfloat[\centering  Free electron-hole dispersion and LO frequency]{{\includegraphics[width= 85mm, height = 67.5mm]{Figures/resolve_crossing.pdf}}}%
%	\hfill
%	\subfloat[\centering Zoomed in plot of dispersion relations]{{\includegraphics[width= 85mm, height = 67.5mm]{Figures/zoom_crossing.pdf}}}%
%	\hfill
	%\subfloat[\centering Lifetime, $\mathcal{T}au$ around 300K]{{\includegraphics[width= 60mm, height = 50mm]{Figures/zoom_tau.pdf}}}%
	%\hfill
%	\caption{Resolving the crossing for exciton-phonon absorption for GaN
%		(a) The variation of the free electron-hole parabolic dispersion for GaN (b) Zoomed in plot showing the crossing - broadening $\eta$ of at least 0.1 meV required. }
%	\label{fig:resolve}
%\end{figure*}
%\begin{table}[ht]&
%\centering
%  \setlength{\mathcal{T}abcolsep}{8pt} % Default is 6pt
%\begin{tabular}{ccccccc}
%\hline
%System &  $E_B$ & $\omega_{LO}$ & $\epsilon_{\infty}$ & $\epsilon_0$ & $m_e$ & $m_h$ \\
%\hline
%GaN & $65$ & $87$ & $5.9$ & $10.8$ & $0.152$ & $1.013$\\
%\hline
%\end{tabular}
%\caption{First principles calculated parameters for GaN from Ref~\cite{alvertis}.}
%\label{table:parameters}
%\end{table}

The phonon mediated exciton--exciton scattering rate for this model system may be derived by making the further parabolic approximation for the exciton bands. The emission linewidth will be exactly zero at zero temperature as the lowest exciton state cannot scatter into a lower energy exciton state via phonon emission. Therefore, the only contribution to the scattering rate will be as a result of the phonon absorption term. Including the temperature dependence of the absorption channel gives the model expression 
\begin{gather}
	\begin{split}~\label{eq:xxfmd}
			\gamma_{\x\x\FMd}(T) &= 2\pi \sum_{\mathbf{q}} \big|G_{SS}(\Gamma,\mathbf{q})\big| ^2 N_{B}(\omega_{LO})\delta\Big(\frac{|\mathbf{q}|^2}{2M}-\omega_{LO}\Big)\\
			&+2\pi \sum_{S'=2}^{N}\sum_{\mathbf{q}} \big|G_{S'S}(\Gamma,\mathbf{q})\big| ^2 N_{B}(\omega_{LO})\\
			&\hspace{15mm}\delta\left(\left[1-\frac{1}{S'^2}\right]E_B+\frac{|\mathbf{q}|^2}{2M}-\omega_{LO}\right) \ ,
	\end{split}
\end{gather}
where $M = m_e+m_h$ is the exciton mass and $N$ is number of exciton bands included in the calculation. The exciton-phonon matrix element for our model system is given by the expression~\cite{cudazzo2020first} 
\begin{gather}
	\begin{split}
		G_{S'S}(\Gamma,\mathbf{q}) = \sum_{\mathbf{k}} \left(A^{S'*}_{\mathbf{k}+\alpha_{e}\mathbf{q}}-A^{S'*}_{\mathbf{k+q}-\alpha_{e}\mathbf{q}}\right)g^{F}_{\mathbf{q}}A^{S}_{\mathbf{k}}\ ,
	\end{split}
\end{gather}
where $\alpha_{e} = \frac{m_e}{M}$ is the weight of the electron coordinate relative to the centre of mass of the exciton.~\cite{toyozawa1958theory,toyozawa1959dynamical,peter2010fundamentals} For the exciton-exciton scattering, we restrict ourselves to scattering between the 1s, 2s, 2p and 3s exciton states in this work, to limit computational effort.

\subsection{Results and discussion}\label{sub:result}

\begin{figure}[ht]
	\centering
	\includegraphics[width= 95mm, height = 79.6mm]{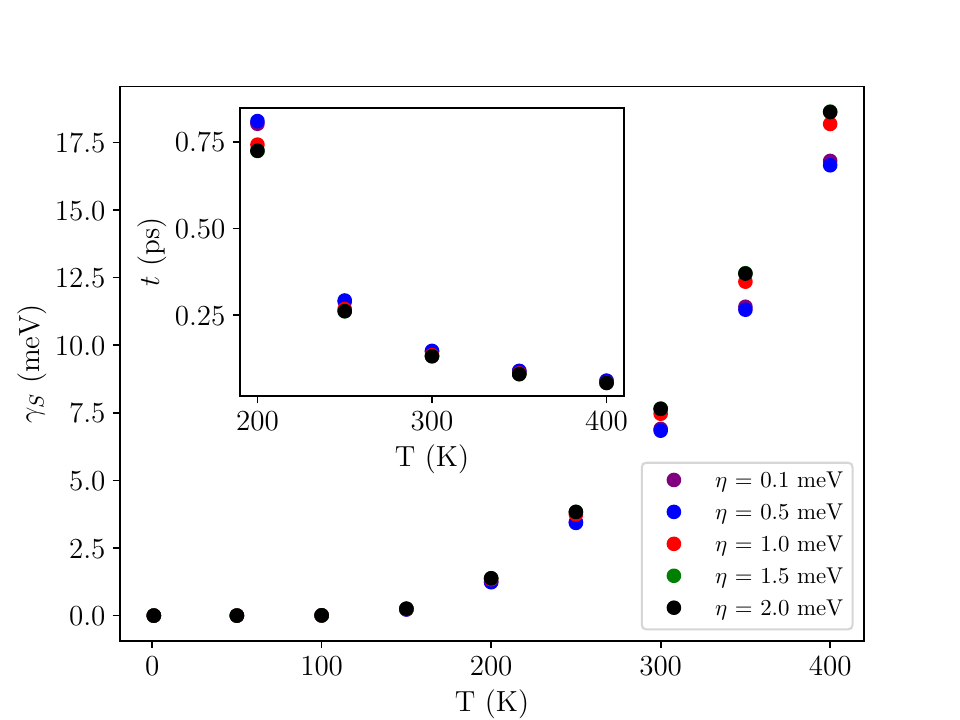}
	\caption{Exciton dissociation lifetime, $\gamma_{\x\eh}$, for GaN model. Main -- variation of the inverse lifetime, Inset -- variation of the lifetime}
	\label{fig:GaN}
\end{figure}
Numerical evaluation of the lifetimes for our model system introduced in Section~\ref{sub:models} requires the knowledge of several material parameters such as the LO phonon frequency and exciton binding energy. The energy conservation condition in the model lifetime expressions for exciton dissociation means they will only become appreciable if the exciton binding energy is less than the LO phonon frequency of the material, $E_{B}<\omega_{LO}$. GaN is one such material that satisfies this criteria,~\cite{xu2002comparative} and in our calculations we use parameters calculated in Ref.~\citenum{alvertis} as follows: $E_B = 65$~meV, $\omega_{LO} = 87$~meV, $\varepsilon_\infty  = 5.9$, $\varepsilon_0 = 10.8$, $m_e = 0.152$ and $m_h = 1.013$.

We numerically approximate the delta functions appearing in all rate expressions by making use of the Kramers-Kronig relation,~\cite{Quantum,mahan2000many} $\delta(x)=\frac{1}{\pi} \I\lim_{\eta\to0^+}\frac{1}{x-i\eta}$, where $\eta$ is a small constant used to numerically resolve the crossing between an exciton that has absorbed a phonon and the resulting free electron-hole pair formed.  We employ the same setup as in Ref.~\citenum{alvertis2023sampling}, using the patched sampling technique, which is necessary for converging the exciton binding energy.~\cite{alvertis2023sampling} We employ $\Gamma$-centered patches, with a cutoff coordinate of $0.07$ or greater (in crystal coordinates), drawn from a fine grid of $100\times100\times100$.~\cite{alvertis} We find that broadenings in the range $0.1-2$~meV are required to resolve the crossing and our model scattering rate to be well converged. 

At zero temperature and for all broadening parameters, the dissociation rate obtained is exactly zero as there are no lower lying states for the exciton to scatter into by phonon emission (Figure~\ref{fig:GaN}). Therefore, at zero temperature, the 1s exciton of GaN has an infinite lifetime. However, as the temperature increases, the phonon absorption channel becomes increasingly present. This is seen in Figure \ref{fig:GaN}, where the dissociation rate and lifetime of the 1s exciton state is plotted as a function of temperature for different broadening values. For the range of broadening parameters plotted ($0.1-2$~meV), we find that the difference between the disociation rates obtained over the temperature range never vary by more than 1 meV. It is clear from Figure \ref{fig:GaN} that the dissociation rate of the exciton is stable with respect to changes in the broadening parameter at all temperatures shown. We find that the lifetimes are well converged for an $\eta$ broadening of $0.5$ meV.

In the inset of Figure~\ref{fig:GaN}, we show the exponential lifetime of the exciton state as a function of temperature for the exciton dissociation channel. At temperatures greater than $0$K, the exciton state acquires a finite lifetime as a result of phonon absorption. The lifetimes obtained are more stable with respect to changes in $\eta$ as the temperature increases due to the inverse relationship between the rate and the lifetime. At 300K, we obtain exciton dissociation lifetimes of $131-146$ fs from our model.

\begin{figure}[ht]
	\centering
	\includegraphics[width= 95mm, height = 79.6mm]{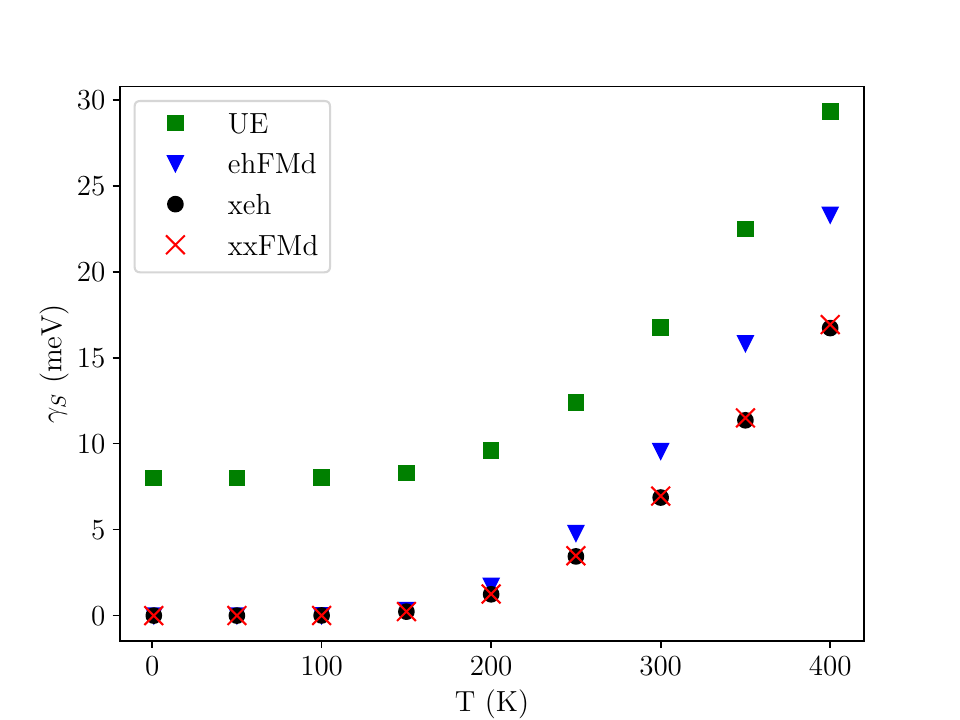}
	\caption{Scattering rate of 1s exciton state for GaN model ($E_{B} = 65$~meV, $\omega_{LO} = 87$~meV) for the different channels outlined in Table~\ref{table:expressions}.}
	\label{fig:lifetimes}
\end{figure}

Figure \ref{fig:lifetimes} shows the model scattering rate of the 1s exciton for GaN parameters for the different scattering channels as described in Table~\ref{table:expressions}. As stated in Section~\ref{sub:ex-ex}, the UE approximation results in an unphysical broadening at zero temperature and gives the fastest scattering rates for the lowest energy exciton. The modified form for the independent electron-hole pair dissociation rate (ehFMd), derived in Eq.~\ref{eq:FM}, is in stark contrast with the UE approximation. Clearly, the ehFMd scattering rate vanishes at zero temperature and only begins to become appreciable at temperatures around $200$ K. This is because the ehFMd rate accounts for exciton dissociation resulting from the single-particle picture of electron-phonon coupling, thereby reducing the lifetime at low temperature compared to the UE approximation result. Intuitively, the ehFMd lifetime accounts for the fact that at low temperature there are insufficient phonons to dissociate the exciton. However, as the temperature increases, the number of phonons increases and the exciton binding energy can be overcome to result in a non-zero dissociation rate.

\begin{figure}[ht]
	\centering
	\includegraphics[width= 95mm, height = 79.6mm]{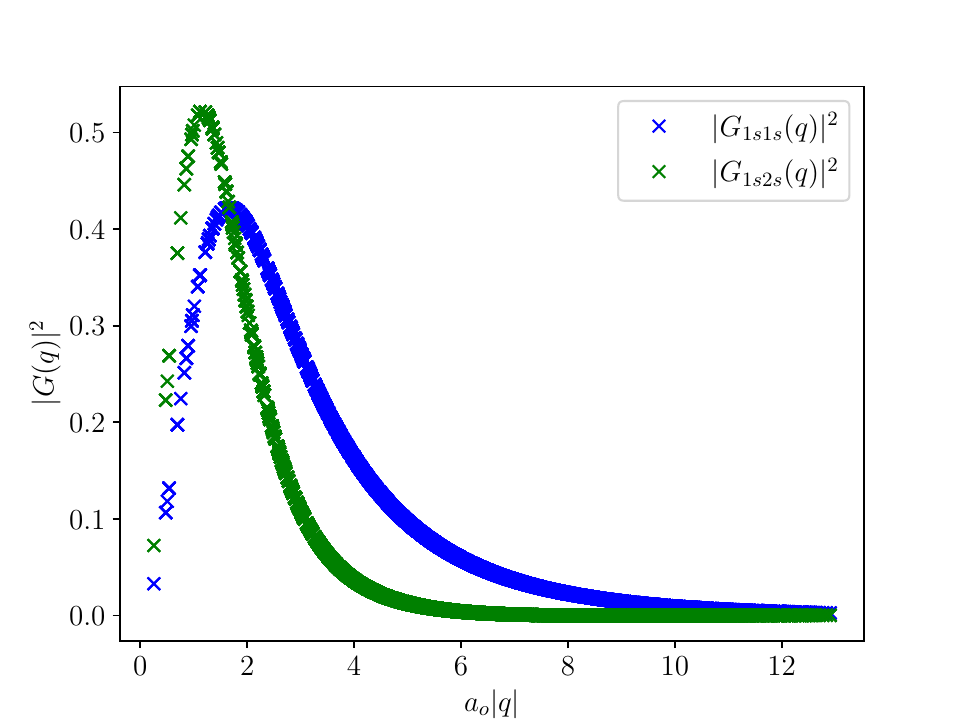}
	\caption{Plot of the model expressions for the 1s-1s and 1s-2s exciton-phonon matrix elements, $|G_{1s1s}(q)|^2$ and $|G_{1s2s}(q)|^2$.}
	\label{fig:ex_gs}
\end{figure}

The same analysis applies for the phonon exchange dissociation rate (xeh). At low temperatures the phonon concentration is not large enough and so the exciton binding energy cannot be overcome. As the temperature increases, the increased concentration of LO phonons leads to rapid absorption by the excitons, resulting in ultrafast dissociation to free electron-hole pairs as the exciton binding energy is undone. The phonon exchange rate shown here for the parameters corresponding to GaN is in good agreement with that obtained from the fully \emph{ab initio} calculation reported in Ref.~\citenum{alvertis}.

In Figure~\ref{fig:lifetimes}, we also plot the exciton--exciton scattering rate obtained with the hydrogenic exciton approximation. This calculation includes the scattering from zero momentum 1s to finite momentum 1s, 2s, 2p and 3s exciton states. 
The xxFMd scattering rate is dominated by 1s-1s and 1s-2s scattering, whilst the 1s-2p and 1s-3s contributions are much smaller. We see a similar scattering rate between the 1s and 2s exciton states relative to the terms describing exciton dissociation to free electron-hole pairs. From Eq.~\ref{eq:xxfmd}, the energy conservation condition for the 1s--2s scattering requires $\omega_{LO} > \frac{3}{4}E_B$. However, the intra- and interband exciton scattering is also strongly dependent on the nature of the exciton-phonon matrix element which differs substantially from that of the Fröhlich vertex. This can be seen in Figure~\ref{fig:ex_gs} where we plot the 1s-1s and 1s-2s exciton-phonon matrix elements as a function of $q$, illustrating nonmonotonic dependence on $q$. As can be seen, the matrix elements vanish for $q = 0$ and increase linearly with $q$ for $qa_o < 1$.~\cite{peter2010fundamentals} In the very large wavevector regime, the matrix elements decay as $\frac{1}{q^{5}}$. Importantly the exciton-phonon vertex vanishes when $m_e=m_h$ due to the cancellation of the effects of the electric field, induced by the electron-hole interaction, on the electron and hole.~\cite{toyozawa1958theory} The resulting balance between the exciton-phonon vertex, large density of exciton states and the energy conservation conditions for 1s-1s and 1s-2s scattering results in a significant rate of exciton--exciton scattering for our model GaN system.

As can be seen in Figure~\ref{fig:lifetimes}, the model exciton-exciton scattering and dissociation rates calculated are very similar. This highlights the importance of accounting for exciton dissociation as well as intra- and interband exciton scattering, particularly when the effective masses of a material are similar, for in this case, the exciton dissociation rate dominates over the exciton-exciton scattering which decreases to zero.

\section{Conclusions and Outlook}\label{sec:conclusion}

In summary, using rearrangement collision theory for exciton--phonon scattering processes, we provide a unified description of scattering between exciton states and exciton dissociation into free charge carriers. This general framework represents an extension of FGR to compute scattering rates between different quasiparticles (eigenstates of different reference Hamiltonians) that reduces to the usual FGR and MBPT results when these eigenstates are of the same Hamiltonian.

From our analysis, we derive two limiting expressions for the rate of phonon-driven exciton dissociation. The first is based on the approximation that the effects of polaron renormalization result in well defined quasiparticle states with infinite lifetime. In this case, the exciton dissociation rate is based solely on the phonon exchange between conduction and valence band states.~\cite{alvertis} This is the explicit correlation of the electron-hole pair due to interaction with the phonon field. The second limiting case occurs whenever the electron or hole band character changes significantly due to the interaction with phonons. This takes place on the `single-particle' level and hence dominates over the correlated phonon exchange between the valence and conduction bands. Importantly, both expressions provide a physically meaningful description of exciton dissociation, enforcing the correct energy conservation. 

We demonstrated all approaches for a model system based on the hydrogenic exciton and Fröhlich approximations to calculate ultrafast temperature-dependent exciton scattering rates for the various scattering channels. Using this model system, we compared different exciton scattering rates from the literature as well as with our recent fully \emph{ab initio} implementation presented in Ref.~\citenum{alvertis}. Table~\ref{table:expressions} provides a summary of the rate expressions derived in this work. 

In polar semiconductors, with electron and hole effective masses with significantly different values, we expect the exciton--exciton scattering channel to be the dominant pathway for materials where the exciton binding energy is much greater than the LO phonon frequency. This is because absorption of phonons can result in ultrafast scattering to a large density of exciton states within the band gap. However, in materials for which the exciton binding energy is actually lower than the LO phonon frequency we expect the exciton dissociation pathway to significantly increase as absorption of phonons takes the exciton to the continuum free charge carrier states. In general, we find that it is necessary to describe both effects due to exciton dissociation and exciton--exciton scattering for a complete picture. Depending on the nature of the `electron'-phonon and `hole'-phonon interactions in a material, the exciton dissociation lifetime may be governed either by a single-particle broadening or a combined `electron-hole'-phonon interaction. However, irrespective of the relative timescale of these processes, our expressions correctly describe phonon mediated exciton dissociation to free charge carriers, a key quantity of interest in optoelectronic devices.

Going beyond the Born approximation to systematically include localization effects resulting from the change in the exciton and conduction/valence band wavefunctions due to interaction with phonons is not straightforward.~\cite{alvertis2023phonon} Potential future directions for including these localization effects in the calculation of exciton scattering rates and band structure renormalization will require a better understanding of polaron interference effects.~\cite{mahanti1972effective}  We hope that this work will provide a foundation for future treatments of general quasiparticle scattering processes in a diverse range of materials.

\vspace{-5mm}
\section*{Acknowledgements}
We acknowledge S. G. Louie for useful comments. C.J.N.C. and M.R.F. acknowledge support from the UK Engineering and Physical Sciences Research Council (EPSRC). J.B.H., A.M.A. and J.B.N. acknowledge the Center for Computational Study of Excited-State Phenomena in Energy Materials (C2SEPEM) as part of the Computational Materials Sciences Program at the Lawrence Berkeley National Laboratory, funded by the U.S. Department of Energy, Office of Science, Basic Energy Sciences, Materials Sciences and Engineering Division, under Contract No. DE-AC02-05CH11231.
%\vspace{5mm}

%\clearpage

\appendix

\section{xxFMd scattering rate}~\label{app:ex_ex}
Inserting Eq.~\ref{eq:ins_s} into Eq.~\ref{eq:matrix}, the $\mathcal{S}$-matrix for the exciton--exciton scattering channel can be expanded as~\cite{lippmann1950variational}
\begin{gather}
    \begin{split}
            \mathbf{S}_{f,i} &= \braket{S',n_\nu+1|S,n_\nu} - 2\pi i \delta(E^{S',\nu}_{f}-E^{S,\nu}_i) \\
            &\times\braket{S',n_\nu+1|V_{ep}\sum_{n=0}^{\infty}\left[G^{+}_0(E^{S,\nu}_i)V_{ep}\right]^{n}|S,n_\nu} \ .
    \end{split}
\end{gather}
Using the orthogonality of the exciton states and the definition of $\ket{\psi^{+}_{i}}$~\cite{lippmann1950variational} as
\begin{equation}
    \ket{\psi^{+}_{i}} = \sum_{n=0}^{\infty}\left[G^{+}_0(E^{S,\nu}_i)V_{ep}\right]^{n}\ket{S,n_\nu}
\end{equation}
the Dyson $\mathcal{S}$-matrix simplifies to 
\begin{equation}
    \mathbf{S}_{f,i} = -2\pi i\ \delta(E^{S',\nu}_f-E^{S,\nu}_i)\braket{S',n_\nu+1|V_{ep}|\psi^{+}_{i}} \ .
\end{equation}
We may also expand the $\mathcal{S}$-matrix to the left using the fact that it is unitary to get~\cite{sunakawa1960theory,sachs1987physics} 
\begin{gather}
\begin{split}
     \mathbf{S}_{f,i} &= -2\pi i \delta(E^{S,\nu}_i-E^{S',\nu}_f)\braket{S',n_\nu+1|V_{ep}|\psi^+_{i}} \\
     &=-2\pi i \delta(E^{S,\nu}_i-E^{S',\nu}_f)\braket{S,n_\nu|V_{ep}|\psi^-_{f}}^* \ ,
\end{split}
\end{gather}
as $\braket{S',n_\nu+1|V_{ep}|\psi^+_{i}}=\braket{S,n_\nu|V_{ep}|\psi^-_{f}}^*$. From the optical theorem, using the Born approximation  $\braket{S',n_\nu+1|V_{ep}|\psi^+_{i}} \approx \braket{S',n_\nu+1|V_{ep}|S,n_\nu}$, and summing over all phonon modes and final excited states $S'$, we find the FGR expression
\begin{equation}~\label{eq:ex_ex_app}
     \gamma_{S} = \Big|2\pi \sum_{S'\nu} \delta(\Omega_S-\Omega_{S'}-\omega_{\nu})\ \big|G_{S'S\nu}\big|^2 \Big|\ , 
\end{equation}
where $G_{S'S\nu}$ is the exciton--phonon vertex.

\section{Modification of the final state}~\label{app:re}
From the initial partitioning of the Hamiltonian as $H=H_0+V_{ep}$, we may write the Green's function as a Dyson series~\cite{sunakawa1960theory,mittleman1961formal,epstein1957theory} 
\begin{equation}
    G^{\pm}(E) = G^{\pm}_0(E) + G^\pm(E)V_{ep}G^{\pm}_0(E)
\end{equation}
where $G^{\pm}_0(E) = \left(E-H_0\pm i\eta\right)^{-1}$ is the reference Green's function corresponding to the reference Hamiltonian. Resolving this identity into the Lippmann-Schwinger equation for the final scattered state 
\begin{equation}
     \ket{\psi^{-}_{2f}} = \ket{vc,n_\nu+1} + G^{-}(E^{vc,\nu}_{f})[V_{eh}+V_{ep}]\ket{vc,n_\nu+1} ,
\end{equation}
we arrive at 
\begin{widetext}
    \begin{gather}
    \begin{split}
        \ket{\psi^{-}_{2f}} &= \ket{vc,n_\nu+1} + G^{-}(E^{vc,\nu}_f)V_{ep}\ket{vc,n_\nu+1} + G^{-}(E^{vc,\nu}_f)V_{eh}\ket{vc,n_\nu+1}\\
        &= \ket{vc,n_\nu+1} + G^{-}(E^{vc,\nu}_f)V_{ep}\ket{vc,n_\nu+1} +\left[G^{-}_0(E^{vc,\nu}_f)+G^{-}(E^{vc,\nu}_f)V_{ep}G^{-}_0(E^{vc,\nu}_f)\right]V_{eh}\ket{vc,n_\nu+1}\\
        &= \left(\ket{vc,n_\nu} + G^{-}_0(E^{vc,\nu}_f)V_{eh}\ket{vc,n_\nu+1}\right) + G^{-}(E^{vc,\nu}_f)V_{ep}\left(\ket{vc,n_\nu+1} + G^{-}_0(E^{vc,\nu}_f)V_{eh}\ket{vc,n_\nu+1}\right) \ .
    \end{split}
\end{gather}
\end{widetext}
By this convenient rearrangement, we can define the renormalized electron-hole pair state as 
\begin{equation}
    \ket{\chi^-_{2f}} = \ket{vc,n_\nu} + G^{-}_0(E^{vc,\nu}_f)V_{eh}\ket{vc,n_\nu+1} \ .
\end{equation}
With this modification, we can write the Lippmann-Schwinger equation for the final scattered state in a convenient form 
\begin{equation}
    \ket{\psi^-_{2f}} = \ket{\chi^-_{2f}} + G^-(E^{vc,\nu}_{f})V_{ep}\ket{\chi^-_{2f}} \ .
\end{equation}

\section{Expansion of the generalized S-matrix}~\label{app:gen}
The modification of the final free electron-hole pair states makes it possible to define a unique $\mathcal{S}$-matrix as the states $\ket{S,n_\nu}$ and $\ket{\chi^{-}_{2f}}$ can be adiabatically connected (via the Gell-Mann and Low theorem~\cite{gell1951bound}) by the same interaction, $V_{ep}$. This generalized $\mathcal{S}$-matrix is given by
\begin{equation}
    \mathbf{S}_{2f;1i} = \braket{\chi^{-}_{2f}|\mathcal{S}(V_{ep})|S,n_\nu} \ .
\end{equation}
The $\mathcal{S}$-matrix can be expanded analogously as
\begin{gather}
    \begin{split}
            &\mathbf{S}_{2f;1i} = \braket{\chi^{-}_{2f}|S,n_\nu} \\
            &- 2\pi i \delta(E^{vc,\nu}_{f}-E^{S,\nu}_i) \braket{\chi^{-}_{2f}|V_{ep}\sum_{n=0}^{\infty}\left[G^{+}(E^{S,\nu}_i)V_{ep}\right]^{n}|S,n_\nu}
    \end{split}
\end{gather}
The initial and final states, $\ket{\psi^{(\pm)}_{2f}}$ and $\ket{\psi^{(\pm)}_{1i}}$, are orthogonal to each other as
\begin{gather}
\begin{split}
        \braket{\psi^{(\pm)}_{2f}|\psi^{(\pm)}_{1i}} &= \braket{\chi^{\pm}_{2f}|U(V_{ep};\mp\infty,0)U(V_{ep};0,\mp\infty)|S,n_\nu} \\
        &= \braket{\chi^{(\pm)}_{2f}|S,n_\nu} = 0 \ . 
\end{split}
\end{gather}
Here, we have used the identity $U(t_0,t_0) = \mathbbm{1}$. This relation holds as the modified states $\ket{\chi^{(\pm)}_{2f}}$ are eigenstates of the same reference Hamiltonian, $H_0$ (with different eigenvalues) and so are orthogonal to $\ket{S,n_\nu}$. Using the orthogonality of the states and the definition of $\ket{\psi^{+}_{1i}}$ gives~\cite{lippmann1950variational}
\begin{equation}\label{eq:s-matrix}
    \mathbf{S}_{2f;1i} = -2\pi i\ \delta(E^{vc,\nu}_f-E^{S,\nu}_i)\braket{\chi^{-}_{2f}|V_{ep}|\psi^{+}_{1i}} \ .
\end{equation}

\section{Generalized optical thorem}~\label{app2}
Performing the derivative in Eq.~\ref{eq:rate_1}, we find
\begin{gather}
    \begin{split}
        w_{2f,1i} &= i\braket{\chi^{-}_{2f}|V_{ep}(t)U(t)|S,n_\nu}\braket{S,n_\nu|U(t)|\chi^{-}_{2f}} \\ 
        &-i\braket{\chi^{-}_{2f}|V_{ep}(t)U(t)|S,n_\nu}^*\braket{S,n_\nu|U(t)|\chi^{-}_{2f}}^*
    \end{split}
\end{gather}
where $U(t)\equiv U(V_{ep};t,-\infty)$ and $^*$ denotes complex conjugation. Using the identity $U(t) = 1 -i\int^{t}_{-\infty} dt' V_{ep}(t')U(t')$, we expand this expression to give 
\begin{gather}
    \begin{split}~\label{eq:rate}
         w_{2f,1i} = &\int^{t}_{-\infty} dt'\braket{\chi^{-}_{2f}|V_{ep}e^{i(E^{vc,\nu}_{f}-H_0)t}U(t)|S,n_\nu} \\
         &\braket{S,n_\nu|V_{ep}e^{i(E^{S,\nu}_{i}-H_0)t'}U(t')|\chi^-_{2f}} \\
         &+\int^{t}_{-\infty} dt'\braket{\chi^{-}_{2f}|V_{ep}e^{i(E^{vc,\nu}_{f}-H_0)t}U(t)|S,n_\nu}^* \\
         &\braket{S,n_\nu|V_{ep}e^{i(E^{S,\nu}_{i}-H_0)t'}U(t')|\chi^-_{2f}}^* 
    \end{split}
\end{gather}
From the Lippmann-Schwinger equations, we have~\cite{lippmann1950variational} 
\begin{subequations}
    \begin{equation}
        e^{-iH_0t}U(t)\ket{S,n_\nu} = e^{-iE^{S,\nu}_{i}t}\ket{\psi^{+}_{1i}}
    \end{equation}
    and
    \begin{equation}
       e^{-iH_0t}U(t)\ket{\chi^{-}_{2f}} = e^{-iE^{vc,\nu}_{f}t}\ket{\psi^{-}_{2f}} ,  
    \end{equation}
\end{subequations}
reducing to 
\begin{gather}
\begin{split}
    w_{2f,1i} &= |\braket{\chi^{-}_{2f}|V_{ep}|\psi^{+}_{1i}}|^2 \int^{t}_{-\infty} dt' \Big[e^{i(E^{vc,\nu}_{f}-E^{S,\nu}_{i})(t-t')} \\
    &+e^{-i(E^{vc,\nu}_{f}-E^{S,\nu}_{i})(t-t')}\Big]
    \\
    &= 2\pi \delta(E^{vc,\nu}_{f}-E^{S,\nu}_{i}) |\braket{\chi^{-}_{2f}|V_{ep}|\psi^{+}_{1i}}|^2 \ .
\end{split}
\end{gather}
We obtain our final expression for the exciton dissociation rate by summing over all final free electron-hole and phonon states $2f\equiv(vc\nu)$ 
\begin{gather}
    \begin{split}
        \gamma_{S} &= \Big|\sum_{2f} w_{2f,1i}\Big| \\
        &=2\pi\Big|\sum_{vc\nu}\delta(E^{vc,\nu}_{f}-E^{S,\nu}_{i})\ |\braket{\chi^-_{2f}|V_{ep}|\psi^+_{1i}}|^2\Big| \ .
    \end{split}
\end{gather}

\vspace{40mm}
%\nocite{*}
%\clearpage
\bibliography{scattering.bib}% Produces the bibliography via BibTeX.

\end{document}